\documentclass[preprint,groupedaddress,pre]{revtex4}
\usepackage{amssymb}
\usepackage{amsmath}

\input{tcilatex}

\begin{document}

\title{Stiff dynamics of electromagnetic two-body motion }
\author{Jayme De Luca}
\email[author's email address:]{ deluca@df.ufscar.br}
\affiliation{Universidade Federal de S\~{a}o Carlos, \\
Departamento de F\'{\i}sica\\
Rodovia Washington Luis, km 235\\
Caixa Postal 676, S\~{a}o Carlos, S\~{a}o Paulo 13565-905}
\date{\today }

\begin{abstract}
We study the stability of circular orbits of the electromagnetic two-body
problem in an electromagnetic setting that includes retarded and advanced
interactions. We give a method to derive the equations of tangent dynamics
about circular orbits up to nonlinear terms and we derive the linearized
equations explicitly. In particular we study the normal modes of the
linearized dynamics that have an arbitrarily large imaginary eigenvalue.
These large imaginary eigenvalues define fast frequencies that introduce a
fast (\emph{stiff}) timescale into the dynamics. As an application of
Dirac's electrodynamics of \emph{point }charges with retarded-only
interactions, we study the conditions for the two charges to perform a fast
gyrating motion of small radius about a circular orbit. The fast gyration
defines an angular momentum of the order of the orbital angular momentum, a
vector that rotates in the orbital plane at a frequency of the order of the
orbital frequency and causes a gyroscopic torque. We explore a consequence
of this multiscale solution, i.e; \ the resonance condition that the angular
momentum of the stiff spinning should rotate exactly at the orbital
frequency. The resonant orbits turn out to have angular momenta that are
integer multiples of Planck's constant to a good approximation. Among the
many qualitative agreements with quantum electrodynamics (QED), the orbital
frequency of the resonant orbits are given by a difference of two
eigenvalues of a linear operator and the emission lines of QED agree with
our predictions within a few percent.
\end{abstract}

\maketitle

\section{Introduction}

We study the Lyapunov stability of quasi-circular orbits of the
electromagnetic two-body problem, a dynamical system with implicitly-defined
delay. The motivation is to understand the balancing of the fast dynamics
described by the delay equations for particle separations in the atomic
magnitude. We give an economical method to derive the linearized equations
of motion about circular orbits. We work in a generalized electromagnetic
setting where the field of the \emph{point} charge is a linear combination
of the advanced and the retarded Li\'{e}nard-Wiechert fields with an
arbitrary coefficient \cite{Dirac}, henceforth called the Eliezer setting
(ES) (see Appendix A). We study in detail a specific feature of the tangent
dynamics; The stiff normal modes of the linearized dynamics, that have an
arbitrarily large imaginary eigenvalue and introduce a fast timescale in the
dynamics. The derivation of the fast normal modes of tangent dynamics is our
main technical contribution to the understanding of this dynamical system.
For some special cases of the ES we find a remarkable quasi-degeneracy of
the tangent dynamics; these cases are Dirac's theory with retarded-only
fields, the action-at-a-distance electrodynamics and the dissipative Fokker
theory of Ref. \cite{dissipaFokker} (see Appendix \ B). Last, we discuss the
dynamics of the hydrogen atom in Dirac's electrodynamics with retarded-only
fields, the special case of the ES of greatest relevance to physics. Having
recognized the existence of the fast dynamics near circular orbits, our
method to find the trajectory starts by balancing the fast dynamics near a
tentative circular orbit. We investigate the conditions for the dynamics to
execute a fast gyrating motion about a circular orbit, henceforth called 
\emph{spinning} and illustrated in Fig. 1. This fast gyration defines an
angular momentum vector of the order of the orbital angular momentum of the
unperturbed circular orbit. This angular momentum of spinning is a vector
that rotates in space with a frequency of the order of the orbital
frequency. We experiment with a typical necessary condition for such
multiscale solution; the resonance condition that the angular momentum of
spinning rotates exactly at the orbital frequency. \ This resonance
condition turns out to be satisfied precisely in the atomic magnitude. We
predict several features of the hydrogen atom of quantum electrodynamics
(QED) \cite{Bohr} with good precision and qualitative detail. The
frequencies of the resonant orbits agree with the lines of QED within a few
percent average deviation. There is also a large body of qualitative
agreement with QED; \ (i) the emitted frequency is given by a difference of
two linear eigenvalues (the Rydberg-Ritz principle) (ii) the resonant orbits
have angular momenta that are approximate multiples of a basic angular
momentum. This basic angular momentum agrees well with Planck's constant and
(iii) the angular momentum of the fast spinning dynamics is of the order of
the orbital angular momentum.

Dirac's 1938 work \cite{Dirac} on the electrodynamics of \emph{point}
charges gave complex delay equations that were seldom studied. Eliezer
generalized Dirac's method of covariant subtraction of infinities in 1947%
\cite{EliezerReview} to include advanced interactions naturally in the
electrodynamics of point charges (the ES) \cite{EliezerReview}. Among the
few dynamics of point charges investigated, another early result of Eliezer 
\cite{Eliezer, Parrott, Andrea, Massimo} revealed a surprising result
(henceforth called Eliezer's theorem); An electron moving in the Coulomb
field of an infinitely massive proton can never fall into the proton by
radiating energy. The result was generalized to motions in arbitrary
attractive potentials \cite{Parrott}, as well as to tridimensional motions
with self-interaction in a Coulomb field \cite{Andrea, Massimo}, finding
that only scattering states are possible. Since our model has the dynamics
of Eliezer's theorem as the infinite-mass limit, a finite mass for the
proton is essential for a physically meaningful dynamics; If the proton has
a finite mass, there is no inertial frame where it rests at all times, and
this in turn causes delay because of the finite speed of light. A finite
mass for the proton is what brings delay into the electromagnetic two-body
dynamics, with its associated fast dynamics. This infinite-mass limit is a
singular limit, because the equations of motion pass from delay equations to
ordinary differential equations! Our understanding of this two-body dynamics
might prove useful for atomic physics and perhaps we can understand QED as
the effective theory of this complex stiff dynamics with delay. We shall
describe the two-body motion in terms of the familiar \emph{center-of-mass
coordinates} and \emph{coordinates of relative separation}, defined as the
familiar coordinate-transformation that maps the two-body Kepler problem
onto the one-body problem with a reduced mass plus a free-moving center of
mass. We stress that in the present relativistic motion the Cartesian
center-of-mass vector is not ignorable, and it represents three extra
coupled degrees-of-freedom. We introduce the concept of resonant dissipation
to exploit this coupling and the many solutions that a delay equation can
have. Resonant dissipation is the condition that both particles decelerate
together, i.e., the center-of-mass vector decelerates, while the coordinates
of relative separation perform an almost-circular orbit, despite of the
energy losses of the metastable center-of-mass dynamics. Last, we stress
that our point charges are not spinning about themselves, but rather
gyrating about a center that is moving with the circular orbit, as
illustrated in Fig. 1, even though we refer to this gyrating motion as \emph{%
spinning}.

The road map for this paper is as follows; In Section IV we give the main
technical part of the paper; We outline our economical method to derive the
tangent dynamics of the circular orbit by expanding the implicit light-cone
condition and the action to quadratic order. This economical method shall be
useful in the further research needed to derive the higher orders and the
complete unfolding of this dynamics. In this Section we also take the stiff
limit of the linear modes of tangent dynamics and introduce the fast
timescale. In Section V we give an application to atomic physics, by
discussing a consequence of balancing the fast dynamics first; This
balancing defines an angular momentum of fast gyration comparable to the
orbital angular momentum, a vector that rotates in space with a slow
frequency of the order of the orbital frequency. In this section we discuss
the heuristic condition that the angular momentum of the fast spinning
motion should rotate at the orbital frequency, a resonance condition that
predicts the correct atomic scales. The earlier sections are a prelude to
Section IV. In Appendix A we give the equations of motion of the
electrodynamics of point charges in the ES. Section II is a review of the
circular orbit solution, to be used in Sections IV and V. In Section III we
give an action formalism for the Li\'{e}nard-Wiechert sector of the ES as a
prelude to the quadratic expansions needed for the linear stability analysis
of Section IV. In Appendix B we derive the general tangent dynamics for
oscillations perpendicular to the orbital plane. The Lemma of resonant
dissipation of Appendix C proves that for the linearized equations of motion
the state of resonant dissipation is impossible. In Appendix D we discuss
how the nonlinear stiff terms balance the leading dissipative term of the
self-interaction force and we estimate the radius of the stiff torus. Last,
in Section VI we put the conclusions and discussion.

\bigskip

\section{The circular orbit solution}

\bigskip

The circular-orbit solution of the isolated electromagnetic two-body problem
of the action-at-a-distance electrodynamics (the ES with $k=-1/2$, see
Appendix A) is used here as the unperturbed orbit\cite{Schonberg,Schild}.
For the ES with $k=-1/2$, the tangent dynamics of the next section is
straightforward Lyapunov stability analysis. In the general case, ( $k\neq
-1/2$), the ES also prescribes a small force opposite to the velocity of
each particle along the circular orbit, such that the circular orbit is not
an exact solution of the equations of motion. The tangent dynamics is then
the starting point of a perturbation scheme to obtain a stiff torus near a
circular orbit (the state of resonant dissipation discussed in Appendix C).

We shall use the index $i=1$ for the electron and $i=2$ for the proton, with
masses $m_{1}$ and $m_{2}$ respectively. We henceforth use units where the
speed of light is $c=1$ and $e_{1}=-e_{2}\equiv -1$ (the electronic charge).
The circular orbit is illustrated in Fig. 2; the two particles move in
concentric circles with the same constant angular speed and along a
diameter. The details of \ this relativistic orbit will be given now; The
constant angular velocity is indicated by $\Omega $ , the distance between
the particles in light-cone is $r_{b}$ and $\theta \equiv \Omega r_{b}$ is
the angle that one particle turns while the light emanating from the other
particle reaches it (the light-cone time lag). The angle $\theta $ is the
natural independent parameter of this relativistic problem. For orbits of
the atomic magnitude the orbital frequency is given to leading order by
Kepler's law%
\begin{equation}
\Omega =\mu \theta ^{3}+...  \label{Kepler}
\end{equation}%
while the light-cone distance $r_{b}$ is given by 
\begin{equation}
r_{b}=\frac{1}{\mu \theta ^{2}}+...  \label{RB}
\end{equation}%
Each particle travels a circular orbit with radius and scalar velocity
defined by 
\begin{eqnarray}
r_{1} &\equiv &b_{1}r_{b},  \label{defradius} \\
r_{2} &\equiv &b_{2}r_{b},  \notag
\end{eqnarray}%
and 
\begin{eqnarray}
v_{1} &=&\Omega r_{1}=\theta b_{1},  \label{defvelocity} \\
v_{2} &=&\Omega r_{2}=\theta b_{2},  \notag
\end{eqnarray}%
for the electron and for the proton, respectively. The condition that the
other particle turns an angle $\theta $ during the light-cone time lag \cite%
{Schild} is 
\begin{equation}
b_{1}^{2}+b_{2}^{2}+2b_{1}b_{2}\cos (\theta )=1,  \label{circularcone}
\end{equation}%
and is henceforth called the unperturbed light-cone condition. As shown in
Appendix B of Ref. \cite{dissipaFokker}, $b_{1\text{ }}$and $b_{2}$ are
approximated by the Keplerian values $b_{1\text{ }}=m_{2}/(m_{1}+m_{2})$ and 
$b_{2\text{ }}=m_{1}/(m_{1}+m_{2})$ \ plus a correction of order $\theta
^{2} $. As discussed in Ref. \cite{Schild}, there is a conserved angular
momentum perpendicular to the orbital plane of the circular orbit given by%
\begin{equation}
l_{z}=\frac{1+v_{1}v_{2}\cos (\theta )}{\theta +v_{1}v_{2}\sin (\theta )},
\label{angular-momentum}
\end{equation}%
where the units of $l_{z}$ are $e^{2}/c$, just that we are using a unit
system where $e^{2}=c=1$. For small values of $\theta $ the angular momentum
of Eq. (\ref{angular-momentum}) is of the order of $l_{z}\sim \theta ^{-1}$.
For orbits of the atomic magnitude, $l_{z}\simeq \theta ^{-1}$ is about one
over the fine-structure constant, $\alpha ^{-1}=137.036$.

\section{\protect\bigskip\ An action for the Lorentz force}

We introduce an action to be used as an economical means to derive the
Lorentz-force sector of the ES equations of motion with the linear
combination of retarded and advanced Li\'{e}nard-Wiechert potentials. In the
following we derive the tangent dynamics by expanding this action to
quadratic order. We henceforth use the dot to indicate the scalar product of
two Cartesian vectors. The Li\'{e}nard-Wiechert action for the Lorentz-force
sector of the ES is 
\begin{equation}
\Theta =-k\int \frac{(1-\mathbf{v}_{1}\cdot \mathbf{v}_{2a})}{r_{12a}(1+%
\mathbf{n}_{12a}\cdot \mathbf{v}_{2a})}dt_{1}+(1+k)\int \frac{(1-\mathbf{v}%
_{1}\cdot \mathbf{v}_{2b})}{r_{12b}(1-\mathbf{n}_{12b}\cdot \mathbf{v}_{2b})}%
dt_{1},  \label{VAintegr}
\end{equation}%
In Eq. (\ref{VAintegr}) $\mathbf{v}_{1}$ stands for the Cartesian velocity
of particle $1$ at time $t_{1}$ and $\mathbf{v}_{2a}$ and $\mathbf{v}_{2b}$
stand for the Cartesian velocities of particle $2$ at the advanced and at
the retarded time $t_{2}$ respectively. The vector $\mathbf{n}_{12a}$ is a
unit vector connecting the advanced position of particle $2$ at time $t_{2}$
to the position of particle $1$ at time $t_{1}$, vector $\mathbf{n}_{12b}$
is a unit vector connecting the retarded position of particle $2$ at time $%
t_{2}$ to the position of particle $1$ at time $t_{1}$ and\textbf{\ }(Below
Eq. (8) of Ref. \cite{dissipaFokker}\ this normal is defined to the opposite
direction, but it was used correctly in Eq. (19) \ of Ref. \cite%
{dissipaFokker} ). Notice that each integral on the right-hand-side of Eq. (%
\ref{VAintegr}) can be cast in the familiar form%
\begin{equation}
\int \frac{(1-\mathbf{v}_{1}\cdot \mathbf{v}_{2c})}{r_{12}(1+\frac{\mathbf{n}%
_{12}\cdot \mathbf{v}_{2c}}{c})}dt_{1}\equiv -\int (V-\mathbf{v}_{1}\cdot 
\mathbf{A})dt,  \label{VAC}
\end{equation}%
where $V$ and $\mathbf{A}$ are the Li\'{e}nard-Wiechert scalar potential and
the Li\'{e}nard-Wiechert vector potential respectively. We have introduced
the quantity $c=\pm 1$ in the denominator of Eq. (\ref{VAC}) such that $c=1$
represents the advanced interaction while $c=-1$ represents the retarded
interaction. The quantities of particle $2$ in Eq. (\ref{VAC}) \ are to be
evaluated at the time $t_{2}$ defined implicitly by 
\begin{equation}
t_{2}=t_{1}+\frac{r_{12}}{c},  \label{lightcone}
\end{equation}%
where $c=\pm 1$ describes the advanced and the retarded light cones,
respectively. Because of this decomposition into $V$ and $\mathbf{A}$ parts,
we henceforth call Eq. (\ref{VAC}) \ the VA interaction. The minimization of
action (\ref{VAintegr}) plus the kinetic energy of particle $1$ yields the
equations of motion for particle $1$ suffering the Lorentz-force of the Li%
\'{e}nard-Wiechert potentials produced by the other particle (see Eq. (\ref%
{Eliezermotion}) ), as shown for example in Ref.\cite{Anderson}).

Besides the Lorentz-force, which is the Lagrangian part of the ES equations
of motion, there is also the dissipative self-interaction force in Eq. (\ref%
{Eliezermotion}). The shortest way to write the ES equations of tangent
dynamics is to add this term to the Lagrangian equations of motion, watching
carefully for the correct multiplicative factor. The stiff limit is
determined by the largest-order derivative appearing in the linearized
equations of motion of Appendix A. In this approximation, the contribution
of the self-interaction force of the ES to the linearized dynamics about a
circular orbit is simply given by the Abraham-Lorentz -Dirac force%
\begin{equation}
\mathbf{F}_{rad}=\frac{2}{3}(1+2k)\mathbf{\dot{a}}\text{,}  \label{LDE}
\end{equation}%
with a renormalized charge. The contribution of the other smaller nonlinear
terms will be given elsewhere (one such nonlinear term is used in the
estimate of Appendix D).

\section{\protect\bigskip Expanding the action to first order: minimization}

\ In the following sections we substitute the circular orbit of Eqs. (\ref%
{defradius}) and (\ref{defvelocity}) plus a planar perturbation into
Eliezer's action (\ref{VAintegr}). We then expand the action up to the
quadratic order to yield the linearized equations of motion. Our economical
derivation of the stability equations anticipates the need to study the
stability of the stiff torus near the circular orbit, a novel solution of
the two-body problem as illustrated in Fig 2. In this work we keep to the
linear stability of circular orbits, which is done with the quadratic
expansion of the action. The variational equations for planar perturbations
are decoupled from the equation for transverse perturbations. We perform the
planar stability analysis using complex gyroscopic coordinates rotating at
the frequency $\Omega $ of the unperturbed circular orbit; The coordinates $%
(x_{j},y_{j})$ of each particle are defined by two complex numbers $\eta
_{j} $ and $\xi _{j}$ ( $j=1$ for electron and $j=2$ for proton) according to%
\begin{eqnarray}
u_{j} &\equiv &x_{j}+iy_{j}\equiv r_{b}\exp (i\Omega t)[d_{j}+2\eta _{j}],
\label{coordinates} \\
u_{j}^{\ast } &\equiv &x_{j}-iy_{j}\equiv r_{b}\exp (-i\Omega t)[d_{j}^{\ast
}+2\xi _{j}],  \notag
\end{eqnarray}%
where $d_{1}\equiv b_{1}$ and $d_{2}\equiv -b_{2}$ are defined in Eqs. (\ref%
{defradius}). Because $x_{j}$ and $y_{j}$ are real, we must have $\eta _{j}$ 
$\equiv \xi _{j}^{\ast }$ but to obtain the variational equations it
suffices to treat $\eta _{j}$ and $\xi _{j}$ as two independent variables in
the Lagrangian. We henceforth call the coordinates of Eq. (\ref{coordinates}%
) the gyroscopic $\eta $ and $\xi $ coordinates. The $d$'s are real numbers
if we choose the origin of times at $t=0$, but we have kept the harmless
star above them to indicate complex conjugation in anticipation to what
follows. Two quantities appear so often in the calculations that we have
named them; (i) The numerator of the action (\ref{VAC}) along circular
orbits, henceforth called $C$, evaluates to 
\begin{equation}
C\equiv 1+b_{1}b_{2}\theta ^{2}\cos (\theta ),  \label{defC}
\end{equation}%
for either retarded ($c=-1$) or advanced interactions ( $c=1$), and\ (ii)
The denominator of action (\ref{VAC}) along circular orbits, divided by $%
r_{b}$ , henceforth called $S$ and defined by

\begin{equation}
S\equiv 1+b_{1}b_{2}\theta \sin (\theta ),  \label{defS}
\end{equation}%
again for either retarded or advanced interactions ($c=\pm 1$). For \ the
stiff limit of Sections V and VI we shall ignore the $O(\theta ^{2})$
corrections and set $C$ $=S=1$. Here we shall derive the electron's equation
of motion only ($j=1$ in Eq. (\ref{coordinates})). The equation for the
proton is completely symmetric and can be obtained by interchanging the
indices. One can derive from Eq. (\ref{coordinates}) that the electron's
velocity at time $t_{1}$ is

\begin{eqnarray}
\dot{u}_{1} &\equiv &v_{1x}+iv_{1y}\equiv \theta \exp (i\Omega
t_{1})[id_{1}+2i(\eta _{1}-i\dot{\eta}_{1}],  \label{velocity1} \\
\dot{u}_{1}^{\ast } &\equiv &v_{1x}-iv_{1y}\equiv \theta \exp (-i\Omega
t_{1})[-id_{1}^{\ast }-2i(\xi _{1}+i\dot{\xi}_{1})].  \notag
\end{eqnarray}%
The velocity of the proton at its time $t_{2}$ can be obtained by simple
interchange of the indices $1$ and $2$, remembering that $d_{1}=b_{1}$ is
defined positive while $d_{2}=-b_{2}$ is defined negative, because at the
same time the particles are in diametrically opposed positions on the exact
circular orbit, such that the exchange operation on the $d^{\prime }s$ is $\
d_{1}\Longleftrightarrow -d_{2}$ (see Fig. 2).

The coordinates of particle $2$ entering in the VA interaction of Eq. (\ref%
{VAC}) are evaluated at a time $t_{2}$ in light-cone with the the present of
particle $1$. Because the implicit light-cone condition has to be expanded
and solved by iteration, it is convenient to define a function $\varphi _{c}$
as 
\begin{equation}
t_{2}\equiv t_{1}+\frac{r_{b}}{c}+\frac{\varphi _{c}}{\Omega }.
\label{light-cone}
\end{equation}%
The above definition is good for both the advanced and the retarded cases, $%
c=1$ defining the advanced light-cone and $c=-1$ defining the retarded
light-cone and the underscore $c$ is to indicate that $\varphi _{c}$ is a
function of $c$. If the perturbation is zero then $\varphi _{c}=0$ and we
are along the original circular orbit, where the light-cone lag is the
constant $r_{b}$. We henceforth call $t$ the present time of particle $1$
(the electron) and we measure the evolution in terms of the scaled-time
parameter $\tau \equiv \Omega t$. The implicit definition of $\varphi _{c}$
by the light-cone condition involves the position of particle $2$ at the
advanced and the retarded time $t_{2}$ as defined by Eq. (\ref{coordinates}),

\begin{eqnarray}
u_{2}(\tau +c\theta +\varphi _{c}) &\equiv &r_{b}\exp (i\Omega
t_{2})[d_{2}+2\eta _{2}(\tau +c\theta +\varphi _{c})],  \label{particle2a} \\
u_{2}^{\ast }(\tau +c\theta +\varphi _{c}) &\equiv &r_{b}\exp (-i\Omega
t_{2})[d_{2}^{\ast }+2\xi _{2}(\tau +c\theta +\varphi _{c})].  \notag
\end{eqnarray}%
as well as the velocity of particle $2$ at the advanced/retarded position

\begin{eqnarray}
\dot{u}_{2}(\tau +c\theta +\varphi _{c}) &\equiv &i\theta \exp (i\Omega
t_{2})[d_{2}+(\eta _{2}-i\dot{\eta}_{2})+2\varphi _{c}(\dot{\eta}_{2}-\ddot{%
\eta}_{2})],  \label{velocity2a} \\
\dot{u}_{2}^{\ast }(\tau +c\theta +\varphi _{c}) &\equiv &-i\theta \exp
(-i\Omega t_{2})[d_{2}^{\ast }+(\xi _{2}+i\dot{\xi}_{2})+2\varphi _{c}(\dot{%
\xi}_{2}+\ddot{\xi}_{2})].  \notag
\end{eqnarray}%
The linear stability analysis involves expanding the equation of motion to
first order in $\eta _{k}$ and $\xi _{k}$ , which in turn is determined by
the quadratic expansion of the action (\ref{VAintegr}) in $\eta _{k}$ and $%
\xi _{k}$. We must therefore carry all expansions up to the second order in
the $\eta \xi $ coordinates. For example the position vector of particle $2$
can be determined to second order by expanding the arguments of $\eta $ and $%
\xi $ in a Taylor series about the value on the unperturbed circular
light-cone for one order only as%
\begin{eqnarray}
u_{2}(\tau +c\theta +\varphi _{c}) &\simeq &r_{b}\exp (i\Omega
t_{2})\{d_{2}+2[\eta _{2}(\tau +c\theta )+\varphi _{c}\dot{\eta}_{2}(\tau
+c\theta )]\},  \label{expansionz2a} \\
u_{2}^{\ast }(\tau +c\theta +\varphi _{c}) &\simeq &r_{b}\exp (-i\Omega
t_{2})\{d_{2}^{\ast }+2[\xi _{2}(\tau +c\theta )+\varphi _{c}\dot{\xi}%
_{2}(\tau +c\theta )]\}.  \notag
\end{eqnarray}%
We shall henceforth indicate the quantities of particle $2$ evaluated
exactly on the light-cone by writing that quantity with a subindex $c$, as
for example $\eta _{2c}\equiv $ $\eta _{2}(\tau +c\theta )$ and $\xi
_{2c}\equiv \xi _{2}(\tau +c\theta )$. In the following we find that $%
\varphi _{c}$ is linear in $\eta $ and $\xi $ to leading order, such that
the next term in the above expansion involves third order terms, which are
not needed for the linear stability analysis of the circular orbit. We shall
always expand quantities evaluated at the perturbed light-cone in a Taylor
series about the \emph{unperturbed} light-cone up to the order needed, such
that our method yields equations with a fixed delay. The perturbed
light-cone is expressed implicitly by the distance from the
advanced/retarded position of particle $2$ to the present position of
particle $1$, described in gyroscopic coordinates by the modulus of the
complex number%
\begin{equation}
u_{c}\equiv u_{1}(\tau )-u_{2}(\tau +c\theta +\varphi _{c}).  \label{deFU}
\end{equation}%
where $c=1$ describes the advanced position and $c=-1$ describes the
retarded position. Using Eq. (\ref{expansionz2a}) this complex number
evaluates to 
\begin{eqnarray}
u_{c} &\equiv &r_{b}\exp (i\Omega t_{2})\{D^{\ast }+2[\exp (-ic\theta
-i\varphi _{c})\eta _{1}(\tau )-\eta _{2c}-\varphi _{c}\dot{\eta}_{2c}]\},
\label{separation} \\
u_{c}^{\ast } &\equiv &r_{b}\exp (-i\Omega t_{2})\{D+2[\exp (ic\theta
+i\varphi _{c})\xi _{1}(\tau )-\xi _{2c}-\varphi _{c}\dot{\xi}_{2c}]\}, 
\notag
\end{eqnarray}%
where we have factored the $\Omega t_{2}$ out and defined the complex number%
\begin{equation}
D\equiv b_{2}+b_{1}\exp (ic\theta +i\varphi _{c}),  \label{defD}
\end{equation}%
that carries an implicit dependence on $\varphi _{c}$ and will later be
expanded to second order as well. Notice that the star appears in the
unusual first line because we are factoring $\Omega t_{2}$ out in Eq. (\ref%
{separation}). At $\varphi _{c}=0$ (the unperturbed circular orbit), the
complex number $D$ defined by Eq. (\ref{defD}) has a unitary modulus. The
last term on the right-hand side of Eq. (\ref{separation}) is a quadratic
form times $\exp (i\Omega t_{2})$ and in the action it only appears
multiplied by a counter-rotating constant term, i.e., a quadratic form times 
$\exp (-i\Omega t_{2})$, such that the product is independent of $t_{2}$. \
Therefore this quadratic term can be integrated by parts and the Gauge term
can be disregarded, as in any Lagrangian. We henceforth call this a
quadratic Gauge simplification and we shall use it in several expressions.
This simplification applied to integrate the terms in $\dot{\eta}_{2c}$ and $%
\dot{\xi}_{2c}$ of Eq. (\ref{separation}) by parts yields%
\begin{eqnarray}
u &\equiv &r_{b}\exp (i\Omega t_{2})\{D^{\ast }+2[(1-i\varphi _{c})\exp
(-ic\theta )\eta _{1}(\tau )-(1-\dot{\varphi}_{c})\eta _{2c}]\},  \label{x}
\\
u^{\ast } &\equiv &r_{b}\exp (-i\Omega t_{2})\{D+2[(1+i\varphi _{c})\exp
(ic\theta )\xi _{1}(\tau )-(1-\dot{\varphi}_{c})\xi _{2c}]\}.  \notag
\end{eqnarray}%
where we have also expanded the exponential of $\varphi $ in the coefficient
of $\eta _{1}$ and \ $\xi _{1}$\ up to the linear order in $\varphi _{c}$,
enough to give the correct quadratic action. Last, the velocity of particle $%
2$ has the following expansion up to a quadratic Gauge%
\begin{eqnarray}
\dot{u}_{2}(\tau +c\theta +\varphi _{c}) &\simeq &i\theta \exp (i\Omega
t_{2})[d_{2}+2(\eta _{2c}-i\dot{\eta}_{2c})(1-\dot{\varphi}_{c})],
\label{velocity2Gauged} \\
\dot{u}_{2}^{\ast }(\tau +c\theta +\varphi _{c}) &\simeq &-i\theta \exp
(-i\Omega t_{2})[d_{2}^{\ast }+2(\xi _{2c}+i\dot{\xi}_{2c})(1-\dot{\varphi}%
_{c})].  \notag
\end{eqnarray}

Using the above quantities, the numerator of the Li\'{e}nard-Wiechert action
(\ref{VAC}) can be calculated as%
\begin{equation}
h_{2}=(1-v_{1}\cdot v_{2c})=(1-\frac{1}{2}\dot{u}_{1}\dot{u}_{2c}^{\ast }-%
\frac{1}{2}\dot{u}_{1}^{\ast }\dot{u}_{2c}).  \label{h2}
\end{equation}%
Last, the denominator of the Li\'{e}nard-Wiechert action (\ref{VAC}) can be
calculated as%
\begin{equation}
h_{4}=r+u\cdot \frac{v_{2a}}{c}=r_{b}(1+\phi )+\frac{z\dot{u}_{2c}^{\ast }}{%
2c}+\frac{z^{\ast }\dot{u}_{2c}}{2c}.  \label{h4}
\end{equation}%
In Eq. (\ref{h4}) we have introduced the scaled delay function $\phi $ by 
\begin{equation}
\varphi _{c}\equiv c\theta \phi .  \label{scaledphi}
\end{equation}%
To relate $\phi $ to the $\xi \eta $ perturbations we expand the implicit
light-cone condition of the perturbed orbit up to the quadratic order%
\begin{equation}
uu^{\ast }=|x_{2}(t+\frac{r_{b}}{c}+\varphi
_{c})-x_{1}(t)|^{2}=c^{2}(t_{2}-t)^{2}=(r_{b}+\frac{\varphi }{\Omega c})^{2}.
\label{quadra}
\end{equation}%
The light-cone condition (\ref{quadra}) is simpler when expressed in terms
of the scaled $\phi $ defined in Eq. (\ref{scaledphi}). The expansion of Eq.
(\ref{quadra}) up to the second order is the following quadratic form in $%
\phi $ 
\begin{eqnarray}
C\phi ^{2}+2S\phi &=&2[(b_{1}\xi _{1}-b_{2}\xi _{2})+(b_{1}\eta
_{1}-b_{2}\eta _{2})]  \label{quadraticform} \\
&&+2[(b_{2}\xi _{1}-b_{1}\eta _{2})\exp (ic\theta )+(b_{2}\eta _{1}-b_{1}\xi
_{2})\exp (-ic\theta )] \\
&&+4[\xi _{1}\eta _{1}-\xi _{1}\eta _{2}\exp (ic\theta )-\xi _{2}\eta
_{1}\exp (-ic\theta )]  \notag \\
&&-2\varphi \lbrack b_{2}(\dot{\xi}_{2}+\dot{\eta}_{2})+b_{1}\dot{\xi}%
_{2}\exp (-ic\theta )+b_{1}\dot{\eta}_{2}\exp (ic\theta )].  \notag
\end{eqnarray}%
The solution to Eq. (\ref{quadraticform}), to first order in the $\xi \eta $
coordinates is given by%
\begin{equation}
S\phi _{(1)}\equiv \lbrack (b_{1}\xi _{1}-b_{2}\xi _{2})+(b_{1}\eta
_{1}-b_{2}\eta _{2})+(b_{2}\xi _{1}-b_{1}\xi _{2})\exp (ic\theta
)+(b_{2}\eta _{1}-b_{1}\eta _{2})\exp (-ic\theta )]  \label{PHI1}
\end{equation}%
As an application of the above, we derive the equations of motion for the
circular orbits of the action-at-a-distance electrodynamics \cite{Schild}
(the ES with $k=-1/2$ ). The equation of motion for $\xi _{1}$ can be
calculated by expanding the VA interaction of Eq. (\ref{VAintegr}) to linear
order 
\begin{equation}
\tilde{\Theta}\equiv 1-\{[\theta ^{2}(S-1)S+C^{2}](b_{1}+b_{2}\cos (\theta
))+S(\theta \sin (\theta )-\theta ^{2}\cos (\theta ))b_{2}\}\frac{(\eta
_{1}+\xi _{1})}{CS^{2}},  \label{linearVA}
\end{equation}%
where the tilde indicates that we scaled $\Theta $ by the factor $\Theta
_{o}=C/(r_{b}S)$ (the value of $\Theta $ along the unperturbed circular
orbit). Scaling the kinetic energy with the same factor and expanding to
first order yields 
\begin{equation}
\tilde{T}_{1}=\frac{-r_{b}m_{1}S}{C\gamma _{1}}+\frac{r_{b}\theta
^{2}m_{1}\gamma _{1}Sb_{1}}{C}(\eta _{1}+\xi _{1}),  \label{linearkinetic}
\end{equation}%
Such that the effective Lagrangian for particle $1$ is%
\begin{equation}
\tilde{L}_{1}\equiv \tilde{T}_{1}+\tilde{\Theta}  \label{action1linear}
\end{equation}%
Since those are linear functions of $\xi _{1}$, the Euler-Lagrange equation
for $\xi _{1}$ is simply $\frac{\partial L_{1}}{\partial \xi _{1}}=0$ or%
\begin{equation}
m_{1}b_{1}r_{b}\gamma _{1}\theta ^{2}S^{3}=[C^{2}+\theta
^{2}S(S-1)](b_{1}+b_{2}\cos (\theta ))+S(\theta \sin (\theta )-\theta
^{2}\cos (\theta ))b_{2}.  \label{Schild3.2}
\end{equation}%
This is Eq. 3.2 of Ref. \cite{Schild}, and the equation for $\eta _{1}$ is
the same condition by symmetry \ (the reason why the circular orbit is a
solution). The equation of motion for particle $2$ can be obtained by
interchanging the indices $1$ and $2$ in Eq. (\ref{Schild3.2}), representing
\ Eq. 3.3 of Ref. \cite{Schild}. In the next section we carry the expansion
of the action to second order, to determine the equations of tangent
dynamics\bigskip .

\section{\protect\bigskip Expanding the action to second order}

The next term of expansion (\ref{linearkinetic}) of the local kinetic energy
of particle $1$ can be calculated with Eq. (\ref{velocity1}) to be%
\begin{eqnarray}
\tilde{T}_{1} &=&T_{o}-\frac{r_{b}Sm_{1}}{C}\sqrt{1-|v_{1}|^{2}}=\frac{%
r_{b}\theta ^{2}m_{1}\gamma _{1}Sb_{1}}{C}(\eta _{1}+\xi _{1})
\label{kinetic1} \\
&&+\frac{Sr_{b}\theta ^{2}m_{1}\gamma _{1}}{2C}[\gamma _{1}^{2}(\xi _{1}+i%
\dot{\xi}_{1}+\eta _{1}-i\dot{\eta}_{1})^{2}-(\xi _{1}+i\dot{\xi}_{1}-\eta
_{1}+i\dot{\eta}_{1})^{2}]+...  \notag
\end{eqnarray}%
This kinetic energy of particle $1$ has the simple quadratic form defined by
two coefficients

\begin{eqnarray}
\tilde{T}_{1} &=&\frac{r_{b}\theta ^{2}m_{1}\gamma _{1}Sb_{1}}{C}(\eta
_{1}+\xi _{1})  \label{quadrat1} \\
&&+M_{1}[\dot{\eta}_{1}\dot{\xi}_{1}+i(\eta _{1}\dot{\xi}_{1}-\xi _{1}\dot{%
\eta}_{1})+\eta _{1}\xi _{1}]+\frac{\theta ^{2}G_{1}}{2}[\xi _{1}^{2}+\eta
_{1}^{2}-\dot{\xi}_{1}^{2}-\dot{\eta}_{1}^{2}]  \notag
\end{eqnarray}%
where $M_{1}\equiv (1+\gamma _{1}^{2})m_{1}\gamma _{1}r_{b}\theta ^{2}S/C$%
and $G_{1}\equiv (\gamma _{1}^{2}-1)m_{1}\gamma _{1}\theta ^{2}S/C$. We also
need the solution of Eq. (\ref{quadraticform}) to second order, $\phi =\phi
_{(1)}+$ $\phi _{(2)}$ , where $\phi _{(1)}$ is given by Eq. (\ref{PHI1})
and $\phi _{(2)}$ is calculated by iteration to be

\begin{eqnarray}
S\phi _{(2)} &=&\varphi (ib_{2}\xi _{1}-ib_{1}\eta _{2}-b_{1}\dot{\eta}%
_{2})\exp (ic\theta )+\varphi (ib_{1}\xi _{2}-ib_{2}\eta _{1}-b_{1}\dot{\xi}%
_{2})\exp (-ic\theta )  \label{PHI2} \\
&&-\varphi b_{2}(\dot{\xi}_{2}+\dot{\eta}_{2})-\frac{1}{2}C[\phi _{(1)}]^{2}
\notag \\
&&+2[\xi _{1}\eta _{1}-\xi _{1}\eta _{2}\exp (ic\theta )-\xi _{2}\eta
_{1}\exp (-ic\theta )]  \notag
\end{eqnarray}%
Next we expand the numerator and the denominator of action (\ref{VAC}), Eqs.
(\ref{h2}) and (\ref{h4}), up to the quadratic order. We use the expansion
of the particle separation, Eq. (\ref{separation}), the velocity of particle 
$1$ (Eq. (\ref{velocity1})), and the expansion of the velocity of particle $%
2 $, Eq. (\ref{velocity2Gauged}). The quadratic expansion of the action is
cumbersome and must be evaluated with a symbolic manipulation software (we
have used Maple 9). \ Integrating by parts and disregarding Gauge terms, we
can bring this quadratic expansion to the following \ " normal form for
particle $1$ "%
\begin{eqnarray}
\Theta _{c} &=&-\frac{i}{2}B(\eta _{1}\dot{\xi}_{1}-\xi _{1}\dot{\eta}%
_{1})+U_{11}\xi _{1}\eta _{1}+\frac{1}{2}N_{11}\xi _{1}^{2}+\frac{1}{2}%
N_{11}^{\ast }\eta _{1}^{2}  \label{quadraVA} \\
&&+R_{c}\xi _{1}\xi _{2c}+R_{c}^{\ast }\eta _{1}\eta _{2c}+P_{c}\xi _{1}\eta
_{2c}+P_{c}^{\ast }\eta _{1}\xi _{2c}+  \notag \\
&&+\frac{Y_{c}}{2}(\xi _{1}\dot{\xi}_{2c}-\xi _{2c}\dot{\xi}_{1})+\frac{%
Y_{c}^{\ast }}{2}(\eta _{1}\dot{\eta}_{2c}-\eta _{2c}\dot{\eta}_{1})+  \notag
\\
&&\frac{\Lambda _{c}}{2}(\xi _{1}\dot{\eta}_{2c}-\eta _{2c}\dot{\xi}_{1})+%
\frac{\Lambda _{c}^{\ast }}{2}(\eta _{1}\dot{\xi}_{1}-\xi _{2}\dot{\eta}%
_{1})+  \notag \\
&&+\frac{T_{c}}{2}\dot{\xi}_{1}\dot{\xi}_{2c}+\frac{T_{c}^{\ast }}{2}\dot{%
\eta}_{1}\dot{\eta}_{2c}+\frac{E_{c}}{2}\dot{\xi}_{1}\dot{\eta}_{2c}+\frac{%
E_{c}^{\ast }}{2}\dot{\eta}_{1}\dot{\xi}_{2c}  \notag
\end{eqnarray}%
Notice that in Eq. (\ref{quadraVA}) the coordinates of particle $2$ appear
evaluated in either the retarded or the advanced unperturbed light-cone,
which is indicated by the subindex $c$. The full quadratic expansion of the
action is composed of the kinetic energy \ plus the partial actions of Eq. (%
\ref{quadraVA}) evaluated at the retarded and the advanced light-cones and
multiplied by the respective Eliezer's coefficient, as in Eq. (\ref{VAintegr}%
) 
\begin{equation}
L_{1}=T_{1}-k\Theta _{+}+(1+k)\Theta _{-}  \label{action1}
\end{equation}%
Notice in Eq. (\ref{quadraVA}) the appearance of the angular-momentum-like
binaries inside parenthesis, a normalization achieved by adding a quadratic
Gauge term to the Taylor series. The coefficient of each normal-form binary
\ is obtained by a simple Gauge-invariant combination of derivatives, for
example the magnetic field coefficient $B$ is given by%
\begin{equation}
B=i[\frac{\partial ^{2}L}{\partial \eta _{1}\partial \dot{\xi}_{1}}-\frac{%
\partial ^{2}L}{\partial \xi _{1}\partial \dot{\eta}_{1}}].  \label{GaugeB}
\end{equation}%
In what follows we evaluate the expansion of the action using a symbolic
software (Maple 9). The linearized Euler-Lagrange equation obtained by
minimizing action (\ref{action1}) respect to the function $\xi _{1}$ is a
linear functional involving the coordinates $\xi _{1}$ , $\eta _{1}$ , $\xi
_{2+}$ , $\eta _{2+}$ , $\xi _{2-}$ and $\eta _{2-}$ as well as their first
and second derivatives%
\begin{eqnarray}
&&l_{1\xi }(\xi _{1},\eta _{1},\xi _{2+},\eta _{2+},\xi _{2-},\eta _{2-})=
\label{EQqsi1} \\
&&-[(N_{11}+\theta ^{2}G)\xi _{1}+\theta ^{2}G\ddot{\xi}_{1}]+[M_{1}(\ddot{%
\eta}_{1}+2i\dot{\eta}_{1}-\eta _{1})-U_{11}\eta _{1}-iB\dot{\eta}_{1}] 
\notag \\
&&-(R_{+}^{\ast }\xi _{2+}+R_{-}^{\ast }\xi _{2-})-(Y_{+}\dot{\xi}_{2+}+Y_{-}%
\dot{\xi}_{2-})+(T_{+}\ddot{\xi}_{2+}+T_{-}\ddot{\xi}_{2-})  \notag \\
&&-(P_{+}\eta _{2+}+P_{-}\eta _{2-})-(\Lambda _{+}\dot{\eta}_{2+}+\Lambda
_{-}\dot{\eta}_{2-})+(E_{+}\ddot{\eta}_{2+}+E_{-}\ddot{\eta}_{2-})  \notag
\end{eqnarray}%
Minimization of action (\ref{action1}) respect to $\eta _{1}$, $\xi _{2}$
and $\eta _{2}$ yields three more linear equations, which together with Eq. (%
\ref{EQqsi1}) compose a system of four linear delay equations. To include
the dissipative self-interaction force into Eq. (\ref{EQqsi1}) we substitute 
$\theta ^{2}=$ $r_{b}^{2}\Omega ^{2}$ into the definition of $M_{1}$ above
Eq. (\ref{PHI2}), such that the Euler-Lagrange equation of the kinetic
energy (\ref{quadrat1}) respect to $\xi _{1}$ can be written as%
\begin{eqnarray}
&&(1+\gamma _{1}^{2})m_{1}\gamma _{1}(S/C)r_{b}^{3}\Omega ^{2}(\ddot{\eta}%
_{1}+2i\dot{\eta}_{1}-\eta _{1})  \label{kinetdP1} \\
&=&r_{b}^{2}(S/C)\frac{d}{dt}(m_{1}\gamma _{1}r_{b}\dot{u}_{1}).  \notag
\end{eqnarray}%
\bigskip On the second line of Eq. (\ref{kinetdP1}) we recognize the
variation of the complex momentum $m_{1}\gamma _{1}r_{b}\dot{u}_{1}$
multiplied by the factor $r_{b}^{2}(S/C)$, \ a complexifyed version of the
planar force. To include the self-interaction into the linearized
Euler-Lagrange equations, we simply add the complexifyed planar force of Eq.
(\ref{LDE}) multiplied by the same factor $r_{b}^{2}(S/C)$ into Eq.(\ref%
{EQqsi1}) 
\begin{equation}
r_{b}^{2}(S/C)\frac{2}{3}(1+2k)r_{b}\frac{d^{3}u_{1}}{dt^{3}}\simeq \frac{%
2\theta ^{3}}{3}(1+2k)\dddot{u}_{1},  \label{LDEqsi1}
\end{equation}%
where the dots represent derivative respect to the scaled time $\tau \equiv
\Omega t$. Using definition (\ref{coordinates}) for the gyroscopic
coordinate $u_{1}$ into Eq. (\ref{LDEqsi1}) yields the offensive force of
Eq. (\ref{offending}) of Appendix C plus the required linearized gyroscopic
version of the triple dots term. The offensive force is a nonhomogeneous
term that will be dealt with in Appendix C. In this section we discard it
and keep only the linear correction to the self force.

Addding the linear term of the self-force to the $\eta _{1}$, $\xi _{2}$ and 
$\eta _{2}$ Euler-Lagrange equations yields a system of four linear delay
equations, which can be solved in general by Laplace transform \cite{Bellman}%
. Here we shall concentrate on the separable normal mode solutions of this
linear system. For that we substitute $\xi _{1}=A\exp (\lambda \tau /\theta
) $, $\eta _{1}=B\exp (\lambda \tau /\theta )$, $\xi _{2}=C\exp (\lambda
\tau /\theta )$ and $\eta _{2}=D\exp (\lambda \tau /\theta )$ into the
linearized equations and we assume that $|\lambda |$ is an order-one or
larger quantity, which is henceforth called the stiff limit (we shall see
that $|\lambda |$ can take values of about $\pi $ or larger while $\theta $
is about $10^{-2}$ in the atomic magnitude).\ The four linear equations in $%
A $, $B$, $C$ and $D$ have a nontrivial solution only if the determinant
vanishes, a condition that we evaluate with a symbolic manipulations
software in the large-$\lambda $ limit (stiff-limit) 
\begin{eqnarray}
&&1-\frac{2(1+2k)\theta ^{2}\lambda }{3}+\frac{(1+2k)\theta ^{4}\lambda ^{2}%
}{9}-\frac{2}{27}\frac{\mu }{M}(1+2k)^{3}\theta ^{6}\lambda ^{3}+...
\label{detXY} \\
&&+\frac{\mu \theta ^{4}}{M}(1+\frac{7}{\lambda ^{2}}+\frac{5}{\lambda ^{4}}%
)[(1+2k)\sinh (2\lambda )-2(1+2k+2k^{2})\cosh ^{2}(\lambda )]  \notag \\
&&-2\frac{\mu \theta ^{4}}{M}(\frac{1}{\lambda }+\frac{5}{\lambda ^{3}}%
)[2(1+2k)\cosh ^{2}(\lambda )-(1+2k+2k^{2})\sinh (2\lambda )]=0  \notag
\end{eqnarray}%
In the special case of $k=-1/2$ Eq. (\ref{detXY}) reduces to Eq. 33 of
reference \cite{dissipaFokker} without the radiative terms, that are ad-hoc
in the dissipative Fokker setting of Ref.\cite{dissipaFokker} (this amounts
to setting $k=0$ in the first line of Eq. (\ref{detXY}) and $k=-1/2$ in the
other two lines). In this work we shall focus on Dirac's retarded-only
electrodynamics of point charges (Eq. (\ref{detXY}) with $k=0$) , with the
following planar-normal-mode condition in the stiff-limit 
\begin{eqnarray}
&&(1+\frac{2}{\lambda _{xy}}+\frac{7}{\lambda _{xy}^{2}}+\frac{10}{\lambda
_{xy}^{3}}-\frac{5}{\lambda _{xy}^{4}}+...)(\frac{\mu \theta ^{4}}{M})\exp
(-2\lambda _{xy})  \label{DiracXY} \\
&=&1-\frac{2}{3}\theta ^{2}\lambda _{xy}+\frac{1}{9}\theta ^{4}\lambda
_{xy}^{2}+...,  \notag
\end{eqnarray}%
Comparing Eq. (\ref{DiracXY}) to Eq. (\ref{DiracZ}) of Appendix B we find
that the quasi-degeneracy of the stiff dynamics exists only for $k=0$,
(retarded-only interactions, Dirac's theory), $k=-1$ (advanced-only
interactions), and $k=-1/2$ (either the dissipative Fokker theory of Ref.%
\cite{dissipaFokker} or the action-at-a-distance electrodynamics).

\section{Stiff torus as a deformed circular orbit}

\bigskip

As discussed in the previous Section, in Appendix B and already explored in
Ref.\cite{dissipaFokker}, there is a remarkable quasi-degeneracy of the
perpendicular and the planar tangent dynamics for some special settings.
Here we shall focus on Dirac's electrodynamics with retarded-only
interactions. Both Eq. (\ref{DiracZ}) and Eq. (\ref{DiracXY}) with $k=-1/2$
in the large-$\lambda $ limit reduce to

\begin{equation}
(\frac{\mu \theta ^{4}}{M})\exp (-2\lambda )=1,  \label{Istar}
\end{equation}%
For hydrogen $(\mu /M)$ is a small factor of about $(1/1824)$. For $\theta $
of the order of the fine structure constant the small parameter $\frac{\mu
\theta ^{4}}{M}\sim 10^{-13}$ multiplying the exponential function on the
left-hand side of Eq. (\ref{Istar}) determines that $\func{Re}(\lambda
)\equiv $ $-\sigma \simeq -\ln (\sqrt{\frac{M}{\mu \theta ^{4}}})$. For the
first $13$ excited states of hydrogen this $\sigma $ is in the interval $%
14.0<|\sigma |<18.0$. The imaginary part of $\lambda $ can be an arbitrarily
large multiple of $\pi $, such that the general solution to Eq. (\ref{Istar}%
) is 
\begin{equation}
\lambda =-\sigma +\pi qi  \label{unperastar}
\end{equation}%
where $i\equiv \sqrt{-1}$ and $q$ is an arbitrary integer. \ Notice that the
real part of $\lambda $ is always negative, such that the tangent dynamics
about the circular orbit is stable in the stiff-limit. The coefficient of $\ 
$the term of order $1/\lambda ^{2}$ is the main\ difference between Eqs. (%
\ref{DiracZ}) and (\ref{DiracXY}). The terms of order $\theta ^{4}\lambda
^{2}$ also differ but they are much less important for $\theta $ in the
atomic range. The exact roots of Eqs. (\ref{DiracXY}) and (\ref{DiracZ})
near the limiting root (\ref{unperastar}) are defined, respectively, by 
\begin{eqnarray}
\lambda _{xy}(\theta ) &\equiv &-\sigma _{xy}+\pi qi+i\epsilon _{1},
\label{pair} \\
\lambda _{z}(\theta ) &\equiv &-\sigma _{z}+\pi qi+i\epsilon _{2},  \notag
\end{eqnarray}%
where $\epsilon _{1}(\theta )$ and $\epsilon _{2}(\theta )$ are real numbers
of the order of $\theta $ and depending on the orbit through $\theta $. The
stiff torus is formed from an initial circular orbit as follows; (i) The
offending force against the velocity dissipates energy and makes the
electron loose radius on a slow timescale (the radiative instability
discussed in Appendix C). (ii) When the electron deviates enough from the
circular orbit, the stiff nonlinear terms compensate the offending terms
(iii) Because of this same deviation from circularity, the stiff terms also
balance the real parts of Eq. (\ref{pair}), the $\sigma ^{\prime }s$ . A
purely harmonic solution to the nonlinear equations of tangent dynamics
appears because the nonlinear stiff terms\ introduce a correction into Eqs. (%
\ref{DiracXY}) and (\ref{DiracZ}), as discussed in Appendix D. The complete
description of this new solution involves expanding the action about the
circular orbit to the fourth order and the coupling to the center-of-mass
recoil and shall be given elsewhere. Because of the large frequency of the
stiff motion, this balancing is established at a very small radius, as
estimated by Eq. (\ref{relativeradius}) of Appendix D. The motion continues
to be stiff because of the imaginary part of the $\lambda ^{\prime }s$ of
Eq. (\ref{pair}). We henceforth assume that this balance is achieved near
the original circular motion. As discussed in Appendix C, \ the circular
orbit recoils in a slow timescale to compensate the momentum loss to the
radiated energy. During the recoil the particle coordinates are described by
a composition of a translation mode and a stiff mode, both with slowly
varying amplitudes, such that the dynamics of the balanced stiff torus is
described by 
\begin{eqnarray}
\xi _{k} &=&A_{k}(T)+u_{k}(T)\exp [(\pi qi+i\epsilon _{1}^{\rho })\Omega
t/\theta ],  \label{multiscalesolution} \\
Z_{k} &=&\func{Re}\{B_{k}(T)+R_{k}(T)\exp [(\pi qi+i\epsilon _{2}^{\rho
})\Omega t/\theta ]\},  \notag
\end{eqnarray}%
where the amplitudes $u_{k}$ and $R_{k}$ must be near the size where the
nonlinearity balances the negative real part of the linear modes. Assuming $%
|u_{1}|\simeq $ $|R_{1}|\simeq \rho $, with $\rho $ given by Eq. (\ref%
{relativeradius}), the angular momentum of the stiff torus along the orbital
plane calculated using (\ref{multiscalesolution}) and disregarding fast
oscillating small terms is%
\begin{eqnarray}
l_{x}+il_{y} &=&\mu r_{b}^{2}\rho ^{2}\frac{\pi q\Omega }{\theta }b_{1}\exp
[i(\epsilon _{2}^{\rho }-\epsilon _{1}^{\rho }+\theta )\Omega t/\theta ]
\label{angularmomentum} \\
&=&\frac{\pi q}{\theta ^{2}}\rho ^{2}\exp [i(\epsilon _{2}^{\rho }-\epsilon
_{1}^{\rho }+\theta )\Omega t/\theta ],  \notag
\end{eqnarray}%
where on the second line of Eq. (\ref{angularmomentum}) we have used formula
(\ref{Kepler}) for $\Omega $ and formula (\ref{RB}) for $r_{b}$. Notice that
we chose the same $q$ on the two perpendicular oscillations in Eq. (\ref%
{multiscalesolution}), such that the fast oscillations beat in the slow
timescale. A typical nonlinear term of this dynamics is the angular momentum
of spin defined by Eq. (\ref{multiscalesolution}). As illustrated in Fig. 1,
the balancing of the fast dynamics, Eq. (\ref{multiscalesolution}), defines
an angular-momentum vector (\ref{angularmomentum}) associated with the fast
spinning. This angular momentum vector rotates at a (\emph{slow!}) frequency
that is generically of the order of the orbital frequency and is determined
solely by the balancing of the fast dynamics. This slowly rotating gyroscope
attached to the electron is the main qualitative feature left after we
balance the stiff delay dynamics. According to Eq. (\ref{angularmomentum})
this slow frequency is%
\begin{equation}
\varpi \equiv \Omega \lbrack 1+\frac{(\epsilon _{2}^{\rho }-\epsilon
_{1}^{\rho })}{\theta }],  \label{rotation}
\end{equation}%
Eq. (\ref{rotation}) defines a frequency of the order of $\Omega $ because $%
\epsilon _{2}^{\rho }$ and $\epsilon _{1}^{\rho }$ are of order $\theta $ by
the definition of expansion (\ref{pair}). The spin angular momentum vector
carries inertia, and its rotation frequency must influence the orbit, such
that the slow trajectory should display some oscillation at the frequency $%
\varpi $ of Eq. (\ref{rotation}). We have assumed that the slow orbit is a
circular orbit of frequency $\Omega $, therefore we must have that $\Omega
=\varpi $ ! (this is the simple physical consequence of solving for the fast
dynamics first!). This physical resonance condition ($\Omega =\varpi $), by
use of Eq. (\ref{rotation}), yields 
\begin{equation}
\epsilon _{1}^{\rho }-\epsilon _{2}^{\rho }=0  \label{resonance1}
\end{equation}%
The calculation of $\epsilon _{1}^{\rho }$ and $\epsilon _{2}^{\rho }$
necessitates and expansion of the action to fourth order and shall be given
elsewhere. We expect nevertheless that the quasi-degeneracy is preserved,
because the stiff limit depends basically on the time light takes to travel
between the particles. Once the stiff torus is near the circular orbit, $%
\epsilon _{1}^{\rho }$ and $\epsilon _{2}^{\rho }$ should differ from $%
\epsilon _{1}$ and $\epsilon _{2}$ by a correction which is essentially the
same because of the quasi-degeneracy plus a correction of order $\theta $%
\begin{equation}
\epsilon _{1}-\epsilon _{2}\simeq \epsilon _{1}^{\rho }-\epsilon _{2}^{\rho
}+b\theta  \label{approximation}
\end{equation}%
Condition (\ref{resonance1}) determines an orbital frequency proportional to
a difference of two eigenvalues, a Rydberg-Ritz-like formula%
\begin{equation}
\Omega =\Omega (\epsilon _{1}-\epsilon _{2})/b\theta =\mu \frac{\theta ^{2}}{%
b}(\epsilon _{1}-\epsilon _{2}).  \label{Rydberg-Ritz}
\end{equation}%
where we have used Kepler's law of Eq. (\ref{Kepler}). The exact calculation
of $b$ demands expanding the action to fourth order and shall be given
elsewhere. \ Since $b$ must be of order one, we shall henceforth set $b=-1$
into Eq. (\ref{approximation}) as a qualitative approximation, yielding 
\begin{equation}
\epsilon _{1}-\epsilon _{2}+\theta =0,  \label{resonance}
\end{equation}%
which has the solutions listed in Table 1. In Ref. \cite{dissipaFokker} we
had postulated heuristically a similar type of resonance for an
electromagnetic-like problem; one involving the real parts of the $\lambda
^{\prime }s$. The resonant orbits found in Ref. \cite{dissipaFokker} were
also in the atomic scale and had the same qualitative behavior found here,
which seems to be a generic feature of these stiff electromagnetic
resonances of the two-body problem. In the last paragraph of Appendix C we
derive this same resonance condition (\ref{resonance1}) from the solvability
of a multiscale asymptotic solution. The above-defined root-searching
problem of Eq. (\ref{resonance}) is well posed and for each integer $q$ it
turns out that one can find a pair of the form (\ref{pair}) if one
sacrifices $\theta $ in Eqs. (\ref{DiracZ}) and (\ref{DiracXY}), i.e., $%
\theta $ must be quantized! According to QED, circular Bohr orbits have
maximal angular momenta and a radiative selection rule ( $\Delta l=\pm 1$)
restricts the decay from level $k+1$ to level $k$ only, i.e., circular
orbits emit the first line of each spectroscopic series (Lyman, Balmer,
Ritz-Paschen, Brackett, etc...), the third column of Table 1. We have solved
Eqs. (\ref{resonance}) and Eqs. (\ref{DiracZ}) and (\ref{DiracXY}) with a
Newton method in the complex $\lambda $-plane. Every angular momentum $%
1/\theta $ determined by Eq. (\ref{resonance}) has the correct atomic
magnitude. These numerically calculated angular momenta $\theta ^{-1}$ are
given in Table 1, along with the orbital frequency in atomic units $%
(137^{3}\Omega )/\mu =137^{3}\theta ^{2}(\epsilon _{2}-\epsilon _{1})$, and
the QED first frequency of each spectroscopic series.

\begin{tabular}{|l|l|l|}
\hline
$l_{z}=\theta ^{-1}$ & $137^{3}\theta ^{2}(\epsilon _{2}-\epsilon _{1})$ & $%
w_{QED}$ \\ \hline
185.99 & 3.996$\times $10$^{-1}$ & 3.750$\times $10$^{-1}$ \\ \hline
307.63 & 8.831$\times $10$^{-1}$ & 6.944$\times $10$^{-2}$ \\ \hline
475.08 & 2.398$\times $10$^{-2}$ & 2.430$\times $10$^{-2}$ \\ \hline
577.99 & 1.331$\times $10$^{-2}$ & 1.125$\times $10$^{-2}$ \\ \hline
694.77 & 7.667$\times $10$^{-3}$ & 6.111$\times $10$^{-3}$ \\ \hline
826.22 & 4.558$\times $10$^{-3}$ & 3.685$\times $10$^{-3}$ \\ \hline
973.12 & 2.790$\times $10$^{-3}$ & 2.406$\times $10$^{-3}$ \\ \hline
1136.27 & 1.752$\times $10$^{-3}$ & 1.640$\times $10$^{-3}$ \\ \hline
1316.44 & 1.127$\times $10$^{-3}$ & 1.173$\times $10$^{-3}$ \\ \hline
1514.40 & 7.403$\times $10$^{-3}$ & 8.678$\times $10$^{-4}$ \\ \hline
1730.93 & 4.958$\times $10$^{-3}$ & 6.600$\times $10$^{-4}$ \\ \hline
1966.77 & 3.379$\times $10$^{-4}$ & 5.136$\times $10$^{-4}$ \\ \hline
2222.70 & 2.341$\times $10$^{-4}$ & 4.076$\times $10$^{-4}$ \\ \hline
\end{tabular}%
\begin{tabular}{|l|}
\hline
$q$ \\ \hline
7 \\ \hline
9 \\ \hline
11 \\ \hline
12 \\ \hline
13 \\ \hline
14 \\ \hline
15 \\ \hline
16 \\ \hline
17 \\ \hline
18 \\ \hline
19 \\ \hline
20 \\ \hline
21 \\ \hline
\end{tabular}

\ 

Caption to Table 1: Numerically calculated angular momenta $l_{z}=\theta
^{-1}$ in units of $e^{2}/c$, orbital frequencies in atomic units $%
(137\theta )^{3}=137^{3}\theta ^{2}(\epsilon _{2}-\epsilon _{1})$, circular
lines of QED in atomic units $w_{QED}\equiv \frac{1}{2}(\frac{1}{k^{2}}-%
\frac{1}{(k+1)^{2}})$ , and the values of the integer $q$ of Eq. (\ref{pair}%
) and Eq. (\ref{multiscalesolution}).

\bigskip

Table 1 illustrates the fact that a resonance involving the tangent dynamics
of a circular orbit predicts magnitudes in the atomic scale, as first
discovered in Ref. \cite{dissipaFokker}. In Ref. \cite{dissipaFokker} we had
to jump the integer $q$ by twenty units\ for a complete quantitative and
qualitative agreement with the Bohr atom. The qualitative agreement achieved
by Table 1 is superior in this way; only for the first two values of $q$
things are not completely natural, and we had to jump some $q^{\prime }s$ ,
after that $q$ increases one by one in qualitative agreement with QED. It
remains to be seen if using the correct $b$ given by the expansion about the
stiff torus will change things quantitatively. The calculation of $b$
necessitates expanding the action to fourth order and shall de done
elsewhere.

\bigskip

\section{Conclusions and Discussion}

\bigskip

In the limit where the proton has an infinite mass, the concept of resonant
dissipation looses meaning because the center-of-mass coordinate no longer
plays a dynamical role. In this singular limit, there is a Lorentz frame
where the proton rests at the origin at all times, and the field at the
electron reduces to a simple Coulomb field in the ES. The two-body dynamics
in the ES reduces then to the dynamical system of Eliezer's theorem;
self-interaction plus a Coulomb field acting on the electron \cite%
{Eliezer,Parrott}. We repeat this correct dynamics because it is very
unpopular \cite{Eliezer,Parrott, Andrea, Massimo}; With inclusion of
self-interaction, it is impossible for the electron to "spiral into the
proton". Neither bound states nor dives are possible, only scattering states
exist. One accomplishment of the present work is to recognize that only the
two-body problem can produce a physically sensible electromagnetic-like
model. By expanding the dynamics about a quasi-planetary orbit (the circular
orbit), we saw the need for a novel ingredient of the electromagnetic
two-body dynamics; the stiff spinning motions!

Another qualitative dynamical picture is suggested by Eliezer's result \cite%
{Eliezer,Parrott}; The dynamical phenomenon that the electron always turns
away from the proton along unidimensional orbits suggests that colinear
orbits are the natural attractors of the dissipative dynamics (the ground
state of the hydrogen atom, with zero angular momentum!). Along such orbits,
the heavy particle (the proton) moves in a non-Coulombian way and the
self-interaction provides the repulsive mechanism that avoids the collision
at the origin. This is again in agreement with the Schroedinger theory,
where the ground state has a zero angular momentum. Again, the infinite-mass
case produces unphysical dynamics; the electron turns away from the proton
but then it runs away \cite{Parrott}. It remains to be researched if the
two-body version of Eliezer's problem has a physical orbit among its
zero-angular momentum orbits, by taking proper care of the delay and its
associated fast dynamics.

The angular momentum vector of fast spinning is rotating at the orbital
frequency, such that the total angular momentum executes precession about an
axis that is determined by the initial condition. Because the angular
momentum of fast spinning is of the order of the orbital angular momentum,
the orbital plane may also oscillate and nutate about an axis that is
determined by the intial condition. The electromagnetic fields generated by
such motion have this axis as a symmetry axis, and therefore this must be
the axis of recoil, since the net radiated momentum calculated with the
Poyting vector must be along the symmetry axis. The unbalanced momentum
along the symmetry axis\ is the precise cause of the recoil. We expect that
our planar circular orbit is a qualitative approximation to this complex
slow motion. The correct multiscale expansion should not start from a
circular orbit, but rather from a general slow motion of both particles plus
a fast perturbation, and then balance the fast dynamics first. The
linearized dynamics about these slowly moving positions is controlled by a $%
6\times 6$ linear matrix containing the stiff delay modes, as the nonlinear
terms couple the $z$ and $xy$ oscillations. This multiscale description
should yield differential equations for the slow guiding orbit by a
Freedholm alternative\cite{Mallet-Paret}, and our preliminary findings
indicate that these complex guiding orbits shall be found in the atomic
region. This detailed research is yet to be done.

In the dynamical process of resonant dissipation, the sharp line is emitted
while the dynamics is locked to the neighborhood of the resonant orbit for a
long timescale. This long timescale is to be compared to the time of
spontaneous decay prescribed by QED for the circular hydrogen lines; about $%
10^{6}$ orbital turns or $10^{-10}$ seconds. This is indeed a very long
timescale compared to the orbital period and should not be confused with the
time for a stiff jump. The stiff modes of Eq. (\ref{pair}) have a frequency
of about $\pi q/\theta \simeq 1000$ times the orbital frequency. This fast
frequency is about $10^{20}$ Hertz and resonates with the X-ray frequencies
used in the Compton effect \cite{Dodd}. This also sets the timescale for a
stiff jump of the dynamics; $10^{-19}$ seconds (10 attoseconds). It is
interesting to notice that this stiff frequency is exactly of the same
magnitude of the zitterbewegung of Dirac's relativistic version of
Schroedinger's equation\cite{Hestenes}. In reference \cite{Hestenes} it is
also proposed that the spin should be associated with a dynamical motion. 

The dynamics starting from an asymptotic resonant orbit to another of a
neighboring $q$ should be described by a stiff jump, as expected generically
from any stiff equation. In Ref \cite{Grasman}, the much simpler Van der Pol
oscillator is worked out in detail as an example of an equation of Lienard
type that exhibits stiff jumps. It is popular to use the concept of\ \ a
quantum jump to describe the stiff passage from one quantum state to
another, but the fact that classical electrodynamics prescribes exactly this
qualitative phenomenon is news. This quasi-instantaneous fast dynamics could
be calculated theoretically and compared to experiment.

The two-body dynamics in the ES solves several conundrums of the hydrogen
atom. Most of those conundrums were created by imagining that the equations
of electrodynamics would accept nonstiff planetary-like orbits. We have seen
that the stiff spinning is necessary if a balance of the fast dynamics is to
be achived in the neighborhood of quasi-planetary orbits, as proved by the
Lemma of resonant dissipation of Appendix C. The balanced stiff spinning is
a novel and non-planetary feature that introduces a gyroscopic torque in the
dynamics. The qualitative agreements with QED are listed in the following;
(i) the resonant orbits are naturally quantized by integers and the radiated
frequencies agree with the Bohr circular lines within a few percent average
deviation. (ii) the angular momenta of the resonant orbits are naturally
quantized with the correct Planck's constant. (iii) the stability analysis
near the balanced stiff torus defines a linear dynamical system with delay,
a dynamical system that needs an initial function as the initial condition,
just like Schroedinger's equation. It remains to be seen if this linear
operator produces a self-adjoint Freedholm alternative\cite{Mallet-Paret},
like in Schroedinger's equation. (iv) The emitted frequencies are given by a
difference of two eigenvalues of this linear operator, like the Rydberg-Ritz
combinatorial principle of quantum physics. (v) the spin angular momentum of
the fast toroidal motion is of the order of the lowest orbital angular
momentum. The spin that we estimate is still a bit too high for QED, where
the electron has a spin angular momentum of $\sqrt{3}\hbar /2$\ , and
perhaps the full equations can cure this\cite{Michael,Tomonaga}. The
calculation of the spin angular momentum for the balanced stiff dynamics
involves a detailed consideration of all stiff terms and is beyond the
estimates of the present work.

We exhibited a new solution of two-body motion in Dirac's electrodynamics of 
\emph{point }charges. The stiff dynamics appears naturally in the two-body
dynamics because of the delay, and it has been so far overlooked. The
balancing of the fast dynamics leads naturally to fast spinning motions; the
multiscale analysis imposes resonances, and these turn out to be satisfied
precisely in the atomic magnitude! The large body of qualitative and
quantitative agreement with QED suggests that further extensive studies of
this two-body dynamics \cite{EliezerReview} could offer an explanation of
QED in terms of a stiff dynamical system with third derivatives and delay.

\section{Acknowledgements:}

I thank Reginaldo Napolitano and Savio B. Rodrigues for discussions.

\bigskip

\section{Appendix A: Eliezer's electrodynamics of point charges}

In Eliezer's generalized electrodynamics\cite{EliezerReview}, the field
produced by the point charge is supposed to be the retarded field plus a
free field $G$%
\begin{equation}
F_{\mu }^{\nu }=F_{\mu ,ret}^{\nu }+G_{\mu }^{\nu }.  \label{eliG}
\end{equation}%
The free field $G$ should satisfy Maxwell's equations, be finite along the
particle's worldline and vanish if the particle is at rest. The choice of
Eliezer in \cite{EliezerReview} is%
\begin{equation}
G_{\mu }^{\nu }=k(F_{\mu ,ret}^{\nu }-F_{\mu ,avd}^{\nu }),  \label{Gk}
\end{equation}%
where $k$ is a fundamental parameter of nature related to the covariant
limit producing the point charge \cite{EliezerReview}. This generalized
electromagnetic setting is henceforth called the Eliezer setting (ES).
Analogously to Dirac's theory \cite{Dirac}, Eliezer's self-interaction is
given by the sourceless combination of half of the retarded Li\'{e}%
nard-Wiechert self-potential minus half of the advanced Li\'{e}nard-Wiechert
self-potential, but multiplied by a renormalizing factor of $(1+2k)$. This
gives the following concise description of the ES; Charges interact with
themselves via the semi-difference of Li\'{e}nard-Wiechert self-potentials
and with other charges via a linear combination of Li\'{e}nard-Wiechert
potentials. In Eliezer's theory the electron and the proton of a hydrogen
atom have the following equations of motion \cite{EliezerReview}%
\begin{eqnarray}
m_{1}\dot{v}_{1\mu }-\frac{2}{3}(1+2k)[\ddot{v}_{1\mu }-||v_{1}||^{2}v_{1\mu
}] &=&-[F_{\mu ,in}^{\nu }+(1+k)F_{\mu 2,ret}^{\nu }-kF_{\mu 2,advt}^{\nu
}]v_{1\nu },  \label{Eliezermotion} \\
m_{2}\dot{v}_{2\mu }-\frac{2}{3}(1+2k)[\ddot{v}_{2\mu }-||v_{2}||^{2}v_{2\mu
}] &=&[F_{\mu ,in}^{\nu }+(1+k)F_{\mu 1,ret}^{\nu }-kF_{\mu 1,adv}^{\nu
}]v_{2\nu },  \notag
\end{eqnarray}%
where $v_{i\mu }$ stands for the quadrivelocity of particle $i$, double bars
stand for the Minkowski scalar product and the dot represents derivative
respect to the proper time of each particle. We are using units where the
electron and the proton have charges $-1$ and $1$ respectively and the speed
of light is $c=1$. The ES has three important limits; (i) For $k=0$ the ES
reduces to Dirac's electrodynamics with retarded-only fields \cite{Dirac}.
(ii) For $k=-1/2$ the ES reduces to the action-at-a-distance equations of
motion derived from Fokker's Lagrangian \cite{FeyWhe} ( notice that the
self-interaction terms disappear). (iii) For $k\simeq $ $-1/2$ the ES
approximates the dissipative Fokker electromagnetic setting of Ref. \cite%
{dissipaFokker} with a charge renormalization controlled by $(1+2k)$. The ES
is discussed in the excellent review of Ref. \cite{EliezerReview}.

\bigskip

\section{\protect\bigskip Appendix B: Linear stability analysis along the $%
\hat{z}$ direction}

In this appendix we perform the linear stability analysis of the circular
orbits for displacements perpendicular to the orbital plane, henceforth
called the $\hat{z}$-direction. We expand to second order the implicit
light-cone condition and the action (\ref{VAintegr}), in the same way of
Section V. The linearized-$z$ dynamics is uncoupled from the planar
dynamics. The Cartesian coordinates of a transversely perturbed circular
orbit are defined as%
\begin{eqnarray}
x_{k}+iy_{k} &\equiv &r_{b}d_{k}\exp (i\Omega t),  \label{Zperturb} \\
x_{k}-iy_{k} &\equiv &r_{b}d_{k}^{\ast }\exp (-i\Omega t),  \notag \\
z_{k} &\equiv &r_{b}Z_{k},  \notag
\end{eqnarray}%
where $k=1$ for the electron and $k=2$ \ for the proton, $Z_{k}$ is the
small transverse perturbation, $d_{1}\equiv b_{1}$ and $d_{2}\equiv -b_{2}$
are defined in Eq. (\ref{defradius}) and $\Omega $ is the orbital frequency
defined above Eq. (\ref{defradius}). We introduce again the delay function $%
\varphi $ of the $Z_{1}$ and $Z_{2}$ perturbations by expanding the
light-cone time $t_{2}$ about the constant lag $r_{b}$ precisely by Eq. (\ref%
{light-cone}). In the following we calculate this homogeneous function $%
\varphi $ of $Z_{1}$ and $Z_{2}$ up to the quadratic order. The distance $%
r_{12}$ entering Eq. (\ref{lightcone}) is to be evaluated from the position
of particle $1$ at time $t_{1}$, to the position of particle $2$ \ at time $%
t_{2}$ defined by Eq.(\ref{Zperturb}). Using $t_{2}$ defined by Eq. (\ref%
{lightcone}) and the orbit of particle $2$ of Eq. (\ref{Zperturb}), the
implicit distance $r_{12}=|t_{2}-t_{1}|$ from particle $1$ at time $t_{1}$
to particle $2$ at time $t_{2}$ is%
\begin{equation}
r_{12}^{2}\equiv r_{b}^{2}(1+\phi
)^{2}=r_{b}^{2}[b_{1}^{2}+b_{2}^{2}+2b_{1}b_{2}\cos (\varphi +\theta
c)]+r_{b}^{2}(Z_{1}-Z_{2c})^{2}.  \label{lightconeZ}
\end{equation}%
Where we again expressed the $\varphi $ of Eq. (\ref{light-cone}) in terms
of the scaled function $\phi $ defined in Eq. (\ref{scaledphi}). Notice that
the $Z$ variations decouple from the planar variations because there is no
mixed linear term of $Z$ times a linear perturbation of the planar
coordinate in Eq. (\ref{lightconeZ}); terms in $Z$ only appear squared.
Expanding Eq. (\ref{lightconeZ}) up to the second order and rearranging
yields

\begin{equation}
\phi ^{2}+2S\phi =(Z_{1}-Z_{2c})^{2}  \label{quadraZ}
\end{equation}%
a quadratic equation for $\phi $ with the regular solution correct to second
order given by%
\begin{equation}
\phi =\frac{1}{2S}(Z_{1}-Z_{2c})^{2}.  \label{PHIZ}
\end{equation}%
The coordinate $Z_{2}$ appears evaluated at the advanced/retarded time in
Eq. (\ref{PHIZ}), and to obtain the action up to quadratic terms it is
sufficient to keep the first term $Z_{2}(\tau _{1}+c\theta +\varphi )$ $%
\simeq Z_{2}(\tau _{1}+c\theta )\equiv Z_{2c}$. Using the $z-$perturbed
orbit defined by Eq. (\ref{Zperturb}) to calculate the numerator of the VA
interaction of Eq. (\ref{VAC}) yields%
\begin{eqnarray}
(1-\mathbf{v}_{1}\cdot \mathbf{v}_{2c}) &=&1+\theta ^{2}\cos (\theta
+c\varphi )b_{1}b_{2}-\theta ^{2}\dot{Z}_{1}\dot{Z}_{2c}\approx  \label{h2Z}
\\
&&C-\theta ^{2}(S-1)\phi -\theta ^{2}\dot{Z}_{1}\dot{Z}_{2c},  \notag
\end{eqnarray}%
and the denominator of the VA interaction of Eq. (\ref{VAC}) is%
\begin{equation}
r_{12}(1+\mathbf{n}_{12c}\cdot \mathbf{v}_{2c}/c)=r_{b}[1+\phi +\theta
cb_{1}b_{2}\sin (\theta c+\varphi _{c})+\theta c(Z_{1}-Z_{2c})\dot{Z}_{2c}].
\label{h4Z}
\end{equation}%
Notice that the quadratic term $Z_{2c}\dot{Z}_{2c}$ on the right-hand side
of Eq. (\ref{h4Z}) can be dropped because it represents an exact Gauge that
does not affect the Euler-Lagrange equations of motion. We also expand the
argument of the sign function of the right-hand side of Eq. (\ref{h4Z})
until the linear term in $\varphi _{c}$, such that the quadratic
approximation to Eq. (\ref{h4Z}) is 
\begin{equation}
r_{12}(1+\mathbf{n}_{12c}\cdot \mathbf{v}_{2c}/c)\approx r_{b}[S+C\phi
+\theta cZ_{1}\dot{Z}_{2c}],  \label{h4ZGauge}
\end{equation}%
where the equivalence sign $\approx $ henceforth means equivalent up to a
Gauge term of second order. Even if a quadratic Gauge term appears in the
denominator, in an expansion up to quadratic order it would still produce a
Gauge and therefore it can be dropped directly from the denominator. In this
way, the expansion up to second order of the VA interaction of Eq. (\ref%
{VAintegr}) is simply%
\begin{eqnarray}
\tilde{\Theta} &\approx &(\frac{C}{r_{b}S})\{(1+k)[1-\theta ^{2}CS^{2}\dot{Z}%
_{1}\dot{Z}_{2-}-\frac{C^{2}S}{2}(Z_{1}-Z_{2-})^{2}+\theta C^{2}SZ_{1}\dot{Z}%
_{2-}]  \label{VAZ} \\
&&-k[1-\theta ^{2}CS^{2}\dot{Z}_{1}\dot{Z}_{2+}-\frac{C^{2}S}{2}%
(Z_{1}-Z_{2+})^{2}-\theta C^{2}SZ_{1}\dot{Z}_{2+}]\}.  \notag
\end{eqnarray}%
Last, we need the kinetic energy along the $z$-perturbed circular orbit,
which \ we express in terms of $Z_{1}$ of definition (\ref{Zperturb}) as%
\begin{equation}
T_{1}=-m_{1}\sqrt{1-v_{1}^{2}}=-\frac{m_{1}}{\gamma _{1}}\sqrt{1-\gamma
_{1}^{2}\theta ^{2}\dot{Z}_{1}^{2}},  \label{kineticZ}
\end{equation}%
where the dot means derivative with respect to the scaled time $\tau $, $%
\gamma _{1}^{-1}\equiv \sqrt{1-v_{1}^{2}}$ , and we have used $\Omega
r_{b}=\theta $. The expansion of Eq. (\ref{kineticZ}) up to second order is%
\begin{equation}
T_{1}=(\frac{1}{r_{b}})\{\frac{-r_{b}m_{1}}{\gamma _{1}}+\frac{\epsilon _{1}%
}{2}\dot{Z}_{1}^{2}+...\},  \label{expaE}
\end{equation}%
where $\epsilon _{1}$ $\equiv m_{1}r_{b}\gamma _{1}\theta
^{2}=r_{b}^{3}m_{1}\gamma _{1}\Omega ^{2}$. The Euler-Lagrange equation of
motion for particle $1$ of the isolated two-body problem is determined from
the quadratic Lagrangian%
\begin{equation}
L_{1}=T_{1}+\tilde{\Theta}.  \label{LagrangeZ1}
\end{equation}%
We shall henceforth disregard the small corrections to the quadratic
coefficients and use $C=1$, $S=1$ , which is the stiff-limit. The equation
of motion for particle $1$ is%
\begin{equation}
\epsilon _{1}\ddot{Z}_{1}=-[Z_{1}-(1+k)Z_{2-}+kZ_{2+}]+\theta \lbrack k\dot{Z%
}_{2+}+(1+k)\dot{Z}_{2-})+\theta ^{2}[k\ddot{Z}_{2+}-(1+k)\ddot{Z}_{2-}].
\label{EQZ1}
\end{equation}%
Notice that the term on the left-hand side of Eq. (\ref{EQZ1}) can be
written as 
\begin{equation}
\epsilon _{1}\ddot{Z}_{1}=r_{b}^{3}m_{1}\gamma _{1}\Omega ^{2}\ddot{Z}%
_{1}=r_{b}^{2}\frac{dp_{z}}{dt},  \label{add1}
\end{equation}%
which is proportional to the force along the $z$-direction multiplied by the
factor $r_{b}^{2}$. According to the equation of motion of the ES (see
appendix A), we must add the following self-interaction term to the
right-hand side of Eq. (\ref{EQZ1})%
\begin{equation}
r_{b}^{2}\mathbf{F}_{rad}=\frac{2}{3}(1+2k)\theta ^{3}\dddot{Z}_{1},
\label{dissi1}
\end{equation}%
where the triple dot means three derivatives with respect to the scaled time
and we have used Eq. (\ref{LDE}). The full linearized equation of motion for 
$Z_{1}$ is%
\begin{eqnarray}
\epsilon _{1}\ddot{Z}_{1} &=&\frac{2}{3}(1+2k)\theta ^{3}\dddot{Z}%
_{1}-[Z_{1}+kZ_{2+}-(1+k)Z_{2-}]  \label{DFEQZ1} \\
&&+\theta \lbrack k\dot{Z}_{2+}+(1+k)\dot{Z}_{2-}]+\theta ^{2}[k\ddot{Z}%
_{2+}-(1+k)\ddot{Z}_{2-}].  \notag
\end{eqnarray}%
The linearized equation for $Z_{2}$ is completely analogous and is obtained
by interchanging $Z_{1\text{ }}$by $Z_{2}$ and $\epsilon _{1}$ by $\epsilon
_{2}$ in Eq. (\ref{DFEQZ1}). (Comparing Eq. (\ref{DFEQZ1}) to Eq. (30) of
Ref. \cite{dissipaFokker} we find that Eqs. (29) and (30) of Ref.\cite%
{dissipaFokker} are both missing a $\theta ^{3}$ factor in front of the $%
\dddot{Z}_{1}$ term, which is just a typo in Ref. \cite{dissipaFokker}
because from Eq. (31) on in Ref \ \cite{dissipaFokker} the self-interaction
force was included correctly). The general solution of a linear delay
equation can be obtained by Laplace transform \cite{Bellman} and is a linear
combination of the following normal mode solutions. A normal mode solution
is obtained by substituting $Z_{1}=A\exp (p\tau )$ and $Z_{2}=B\exp (p\tau )$
into the two linearized equations, and requires the vanishing of the
following $2\times 2$ determinant%
\begin{equation}
\det Z\equiv \left\vert 
\begin{array}{cc}
1+\epsilon _{1}p^{2}-\frac{2}{3}(1+2k)\theta ^{3}p^{3} & G(\theta ,p) \\ 
G(\theta ,p) & 1+\epsilon _{2}p^{2}-\frac{2}{3}(1+2k)\theta ^{3}p^{3}%
\end{array}%
\right\vert ,  \label{detZ}
\end{equation}%
where $G(\theta ,p)\equiv \lbrack 1-(1+2k)p\theta -\theta ^{2}p^{2}]\cosh
(p\theta )-[(1+2k)-\theta p-(1+2k)\theta ^{2}p^{2}]\sinh (p\theta )$. The
stiff limit obtained when $p\theta $ is large, such that the hyperbolic
functions of the $G(\theta ,p)$ acquire a large magnitude \cite{astar2B}. In
the following we use the Coulombian limit values, $b_{1}=m_{2}/M$ and $%
b_{2}=m_{1}/M$ , (see Appendix B of Ref. \cite{dissipaFokker}) to evaluate
determinant (\ref{detZ}), such that

\begin{eqnarray}
\epsilon _{1} &=&\frac{M}{m_{2}},  \label{zeroth} \\
\epsilon _{2} &=&\frac{M}{m_{1}}.  \notag
\end{eqnarray}%
For small $\theta $, the second-order and higher even-order corrections to\
Eq. (\ref{zeroth}) are very small. Defining $p\equiv \lambda /\theta $ and
using Eq. (\ref{zeroth}) we obtain%
\begin{eqnarray}
\frac{\mu \theta ^{4}}{M\lambda ^{4}}(\det Z) &=&1-\frac{2}{3}(1+2k)\theta
^{2}\lambda +\frac{4}{9}\frac{\mu }{M}\theta ^{4}\lambda ^{2}  \label{EQ5Z}
\\
&&-\frac{\mu \theta ^{4}}{M}\{[1-\frac{1}{\lambda ^{2}}+\frac{(1+2k)}{%
\lambda }]\cosh (\lambda )-[\frac{1}{\lambda }+(1+2k)(1-\frac{1}{\lambda ^{2}%
})]\sinh (\lambda )\}^{2},  \notag
\end{eqnarray}%
where we have dropped small $O(\theta ^{2})$ terms. Equation (\ref{EQ5Z})
exhibits the generic feature of the stiff-limit; that the hyperbolic
functions always appear multiplied by the very small coefficient $\mu \theta
^{4}/M$ . The stiff-mode condition defined by Eq. (\ref{EQ5Z}) ($\det Z=0$)
with $k=-1/2$ is equation (33) of Ref.\cite{dissipaFokker}, i.e.,%
\begin{equation}
1-\frac{2}{3}\theta ^{2}\lambda +\frac{4\mu }{9M}\theta ^{4}\lambda ^{2}-%
\frac{\mu \theta ^{4}}{M}[(1-\frac{1}{\lambda ^{2}})[\cosh ^{2}(\lambda )-%
\frac{1}{\lambda }\sinh (2\lambda )]^{2}=0.  \label{fourthZ}
\end{equation}%
Notice that in Ref \cite{dissipaFokker} there is a typo in passing from Eq.
(33) to Eq. (34); Eq. (34) is missing a bracket that should start after the $%
\frac{\mu \theta ^{4}}{M}$ factor and close at the end of Eq. (34). The
special case of $k=0$ is Dirac's theory with retarded-only fields%
\begin{equation}
1-\frac{2}{3}\theta ^{2}\lambda +\frac{4\mu }{9M}\theta ^{4}\lambda ^{2}-%
\frac{\mu \theta ^{4}}{M}[1+\exp (-2\lambda )](1+\frac{2}{\lambda }-\frac{1}{%
\lambda ^{2}}-\frac{1}{\lambda ^{3}}+\frac{1}{\lambda ^{4}})=0,
\label{DiracZ}
\end{equation}%
where the appearance of the negative exponential only is related to the
retardation-only, instead of the hyperbolic functions related to advanced
and retarded interactions. Comparing Eq. (\ref{EQ5Z}) to Eq. (\ref{detXY})
we can see that the phenomenon of quasi-degeneracy exists only for $k=0$
(Dirac's theory) and for $k=-1/2$ (the action-at-a-distance electrodynamics
and the dissipative Fokker theory of Ref. \cite{dissipaFokker}).

\bigskip

\section{\protect\bigskip Appendix C: Lemma of resonant dissipation}

\bigskip

Substituting the circular orbit into the equations of motion, Eqs. (\ref%
{Eliezermotion}), one finds at leading order for the electronic motion an
offending term described by a force opposite to the electronic velocity%
\begin{equation}
r_{b}^{2}F_{1}=r_{b}^{2}\frac{2}{3}\dddot{x}_{1}=-\frac{2e^{2}}{3c^{3}}%
\Omega ^{2}r_{b}^{2}\dot{x}_{1}=-\frac{2}{3}\theta ^{3}.  \label{offending}
\end{equation}%
where we have included the scaling factor $r_{b}^{2}$ as of Eq. (\ref{dissi1}%
). Along the trajectory of the proton, the delayed Li\`{e}nard-Wiechert
interaction with the electron is the main offending force against the
velocity, instead of the much smaller protonic self-interaction. Using the
Page series in the same way of Ref. \cite{PRL}, we find a force against the
protonic velocity of the same magnitude of Eq.(\ref{offending}), as
illustrated in Fig. 2. These offending forces destabilize the unperturbed
circular motion with a small torque and a slow dissipation. In quantum
electrodynamics (QED) the circular Bohr orbits\cite{Bohr} of hydrogen
correspond to excited states that decay to the ground state in a life-time
of about $10^{6}$ turns. Using Eq. (\ref{offending}) to estimate the energy
dissipation along circular orbits of the atomic magnitude, one finds a net
dissipation of about $4$ electron-volts after the life-time of $10^{6}$
turns, an order-one fraction of the biding energy ($13.6$ electron-volts).
This radiative instability in the slow timescale is an incomplete picture of
the dynamics because it disregards the fast stiff motion that is present
because of the delay. Let us postulate heuristically that along some special
circular orbits the fast dynamics encircles the circular orbit as
illustrated in Fig. 1. If the linearization about the circular orbit, Eq. (%
\ref{multiscalesolution}), is averaged over a timescale of some turns, the
stiff dynamics goes away and the resulting variable is a slow moving drift
(representing the recoil of the state of resonant dissipation). For that
slow motion we have%
\begin{eqnarray}
\bar{\xi}_{1}(\tau ) &=&\bar{\xi}_{1+}(\tau )=\bar{\xi}_{1-}(\tau ),
\label{averagerecoil} \\
\bar{\xi}_{2}(\tau ) &=&\bar{\xi}_{2+}(\tau )=\bar{\xi}_{2-}(\tau ),  \notag
\end{eqnarray}%
since we are averaging the argument over a time of many turns, a time much
longer than the delay $\theta $. We are ready to prove that the dynamics of
the state of resonant dissipation must necessarily involve a nonlinear term
of the expansion about a circular orbit.

\emph{Lemma}: If the averaged dynamics is determined only by the linear
terms plus the offending forcing, the state of resonant dissipation is
impossible. To show this we average the full equation of motion of the $\xi
_{1}$ variable, which is 
\begin{equation}
\left\langle \frac{2\theta ^{3}}{3}(\dddot{u}_{1}+ib_{1})\right\rangle +\bar{%
l}_{1\xi }-ib_{1}\theta ^{3}+NL=0,  \label{nonlinearbalance}
\end{equation}%
where the term inside brackets is the average of the linearized correction
to the self-interaction force, $\bar{l}_{1\xi }$ stands for the average of
Eq. (\ref{EQqsi1}) and $NL$ stands for the average of the nonlinear terms.
The time averages of the term inside brackets and of the first and second
derivatives in Eq. (\ref{EQqsi1}) are zero, such that discarding the
nonlinear term of Eq. (\ref{nonlinearbalance}) yields the \emph{%
non-homogeneous} linear equation%
\begin{eqnarray}
&&-(N_{11}+\theta ^{2}G)\bar{\xi}_{1}-(M_{1}+U_{11})\bar{\eta}_{1}
\label{nonlinearlemma} \\
&&-(R_{+}^{\ast }+R_{-}^{\ast })\bar{\xi}_{2}-(P_{+}+P_{-})\bar{\eta}_{2} 
\notag \\
&=&ib_{1}\theta ^{3}.  \notag
\end{eqnarray}%
Performing the same average for the other three coordinates, we obtain a
system of four non-homogeneous linear equations. The matrix of this linear
system has a zero determinant that is directly related to the Galilean
translation mode of the circular orbit, such that it defines a singular
linear operator. Moreover, the forcing term turns out to be out of the image
of the singular linear operator, such that the linear system has no solution
at all! This completes the proof of our lemma.

Last, we mention a derivation of Eq. (\ref{resonance1}) from the equations
of motion in the form (\ref{nonlinearbalance}); This can be accomplished by
placing the linearization about the stiff torus on the left-hand side
(henceforth called the Kernel) and the forcing and nonlinear terms on the
right-hand side. In this way the right-hand side has to be orthogonal to the
left-eigenvectors of the Kernel, which is the Freedholm alternative theorem
discussed for example in Refs. \cite{vereador,Mallet-Paret}. Multiplying the
Kernel by a left-eigenvector and integrating on the fast timescale has
exactly the form of the conservation law for angular momentum on the slow
time scale. This shall be studied elsewhere.

\section{\protect\bigskip Appendix D: Harmonic solution and spinorial
unfolding}

The lemma of resonant dissipation of Appendix C suggests the participation
of the nonlinear terms to achieve the state of resonant dissipation of Fig.
1. We shall see in the following that because of the fast frequency in the
stiff terms, even at small amplitudes the nonlinear stiff terms give a large
contribution to the self-force. Our perturbation scheme started from a
guiding circular orbit that is not a solution to the equations of motion,
because of the offending force of Eq. (\ref{offending}). In the following we
show how to balance this offending force at a small distance away from the
circular orbit. The full Lorentz-Dirac self-force is given in page 116 of
Ref \cite{Rohrlich}, expressed in terms of the Cartesian velocity of the
particle, $\mathbf{v}_{1}$, its acceleration $\mathbf{a}_{1}$ and its second
acceleration $\mathbf{\dot{a}}_{1}$ respectively. This full self-force
multiplied by the convenient scaling factor of Eq. (\ref{dissi1}) is%
\begin{equation}
r_{b}^{2}\mathbf{F}_{1}=\frac{2}{3}\gamma _{1}^{3}r_{b}^{2}\{\mathbf{\dot{a}}%
_{1}+3\gamma ^{2}(\mathbf{v}_{1}\cdot \mathbf{a}_{1})\mathbf{a}_{1}+\gamma
_{1}^{2}[(\mathbf{v}_{1}\cdot \mathbf{\dot{a}}_{1})+3\gamma _{1}^{2}(\mathbf{%
v}_{1}\cdot \mathbf{a}_{1})^{2}]\mathbf{v}_{1}\}.  \label{nonlidissi1}
\end{equation}%
Assuming that the particle coordinates are described by a slow term plus a
fast term, Eq. (\ref{multiscalesolution}), we use Eq. (\ref{nonlidissi1}) as
follows; We estimate each coefficient of the $\mathbf{v}_{1}$, $\mathbf{a}%
_{1}$ and $\mathbf{\dot{a}}_{1}$ terms by substituting\ the fast component
of the trajectory, Eq. (\ref{multiscalesolution}), and taking the time
average. For the estimate below, the vectors $\mathbf{v}_{1}$, $\mathbf{a}%
_{1}$ and $\mathbf{\dot{a}}_{1}$ are replaced by the slow quantities
evaluated along the unperturbed orbit. For example the coefficient of the
velocity term in Eq. (\ref{nonlidissi1}) ($r_{b}^{2}$ times the term inside
the square brackets on the right-hand side of Eq. (\ref{nonlidissi1}) ) is%
\begin{equation}
r_{b}^{2}\gamma _{1}^{2}[(\mathbf{v}_{1}\cdot \mathbf{\dot{a}}_{1})+3\gamma
_{1}^{2}(\mathbf{v}_{1}\cdot \mathbf{a}_{1})^{2}]=\gamma _{1}^{2}|\lambda
|^{4}\rho ^{2}(3\gamma _{1}^{2}|\lambda |^{2}\rho ^{2}-1)  \label{coefv1}
\end{equation}%
Notice that for a large enough $\rho $ there is a critical value where this
coefficient changes sign, or 
\begin{equation}
3\gamma _{1}^{2}|\lambda |^{2}\rho ^{2}>1,  \label{estimate1}
\end{equation}%
We show in the following that the value of $\rho _{1}$ must be very near
this critical value. The self-forces for the electron and proton, both
measured along the direction of the electronic velocity, are

\begin{eqnarray}
r_{b}^{2}F_{1} &=&\frac{2}{3}[-\theta ^{3}b_{1}+\gamma _{1}^{2}|\lambda
|^{4}\rho _{1}^{2}(3\gamma _{1}^{2}|\lambda |^{2}\rho _{1}^{2}-1)\theta
b_{1}],  \label{balance} \\
r_{b}^{2}F_{2} &=&\frac{2}{3}[\theta ^{3}b_{2}-\gamma _{2}^{2}|\lambda
|^{4}\rho _{2}^{2}(3\gamma _{2}^{2}|\lambda |^{2}\rho _{2}^{2}-1)\theta
b_{2}],
\end{eqnarray}%
where $b_{1}$ and $b_{2}$ are given by Eq. (\ref{defradius}). In the state
of resonant dissipation, the relative separation of the particles executes a
circular motion, such that the tangential force must vanish%
\begin{equation}
b_{1}F_{1}-b_{2}F_{2}=0,  \label{force}
\end{equation}%
with $b_{1\text{ }}=m_{2}/(m_{1}+m_{2})$ and $b_{2\text{ }%
}=m_{1}/(m_{1}+m_{2})$. Because $b_{2}\ll 1$ we can disregard the protonic
contribution such that Eq. (\ref{force}) yields%
\begin{equation}
\rho _{1}=\sqrt{\frac{1}{3|\lambda |^{2}}(1+\frac{\theta ^{2}}{|\lambda
|^{4}\rho _{1}^{2}})}\simeq \frac{1}{\sqrt{3|\lambda |^{2}}}.
\label{relativeradius}
\end{equation}%
This distance is a few percent of the orbital radius, and it is some $500$
classical electronic radia for $\theta $ \ in the atomic scale. The spin
angular momentum of the electron calculated by Eq. (\ref{angularmomentum})
is \ $|l|\sim 1/(6\pi q\theta ^{2})$, which is of the order of the orbital
angular momentum.

The nonlinear correction to Eq. (\ref{DiracZ}) can be estimated by taking
into account the nonlinear terms of the Lorentz-Dirac self-force. This
correction to Eq. (\ref{DiracZ}), estimated along a \emph{harmonic} solution
for the averaged equations of tangent dynamics, in a way explained above,
yields 
\begin{eqnarray}
&&(1+\frac{2}{\lambda _{z}}-\frac{1}{\lambda _{z}^{2}}-\frac{1}{\lambda
_{z}^{3}}+\frac{1}{\lambda _{z}^{4}}+...)(\frac{\mu \theta ^{4}}{M})\exp
(-2\lambda _{z})  \label{DiracZa} \\
&=&1-2\rho ^{2}|\lambda |^{3}-\frac{2}{3}\theta ^{2}\lambda _{z}+\frac{4\mu 
}{M}\theta ^{4}\lambda _{z}^{2}+...,  \notag
\end{eqnarray}%
Because of the extra term on the right-hand side of Eq. (\ref{DiracZa}), a
harmonic solution to the \emph{nonlinear }equations of tangent dynamics
exists if%
\begin{equation}
\rho \simeq \frac{1}{\sqrt{2}|\lambda |^{3/2}}.  \label{evala}
\end{equation}

The fact that the two \emph{necessary }and \emph{independent} estimates (\ref%
{relativeradius}) and (\ref{evala}) agree suggests that the stiff torus of
Fig. 1 is a solution of Dirac's equations of motion\cite{Dirac} for the
two-body problem! In summary, our perturbative scheme shows that the stiff
torus of Fig. 2 is a solution to the equations of motion as follows; (i) We
expand about a circular orbit and calculate the radius of the stiff torus by
the condition that a \emph{harmonic} solution to the nonlinear equations of
tangent dynamics exists (Eq. \ref{evala}). (ii) This balanced stiff spinning
about the circular orbit cancels the offending force of Eq. (\ref{offending}%
), predicting a radius by (\ref{relativeradius}) that is in agreement with
the radius of (\ref{evala}). (iii) Resonance condition (\ref{resonance1}) is
a necessary condition for the existence of an asymptotic expansion of the
dynamics about the stiff torus, i.e., the condition for a smooth recoil of
the atom as a \emph{bound state}. The detailed unfolding of the equations of
tangent dynamics is beyond the estimates of this Appendix and necessitates
the full nonlinear equations.

\section{Captions}

\bigskip

Fig. 1: The guiding-center circular orbit of each particle is illustrated by
dashed lines. Particle trajectories are the stiff tori gyrating about the
guiding-center circular orbit of each particle (dark solid lines).
Illustrative purposes only, the radia are not on scale. Arbitrary units.

\bigskip

\bigskip

Fig. 2: \ The unperturbed circular orbit with the particles in diametral
opposition at the same time in the inertial frame. Indicated is also the
advanced position of particle $1$ and the angle travelled during the
light-cone time. The drawing is not on scale; The circular orbit of the
proton has an exaggerated radius for illustrative purposes. Arbitrary units.


\bigskip

\begin{references}

\bibitem{Dirac} P. A. M.Dirac, {\it Proceedings of the Royal Society of London, ser. A}
{\bf 167},148 (1938).

\bibitem{dissipaFokker} J. De Luca, { \it Physical Review E }, {\bf 71} 056210 (2005).

\bibitem{Bohr} N. Bohr, Philos. Mag., {\bf 26}, 1 (1913);  {\bf 26}, 476 (1913).

\bibitem{EliezerReview}C. Jayaratnam Eliezer, {\it Reviews of Modern Physics}, {\bf 19} (1947). 

\bibitem{Eliezer} C.J. Eliezer, {\it Proc. Cambridge Philos. Soc.} {\bf 39}, 173 (1943).

\bibitem{Parrott} S. Parrott, {\it Foundations of Physics} {\bf 23}, 1093 (1993).


\bibitem{Andrea}A. Carati, {\it  J. Phys. A: Math. Gen.} {\bf 34}, 5937 (2001).


\bibitem{Massimo}M. Marino, {\it  J. Phys. A: Math. Gen } {\bf 36}, 11247 (2003). 

\bibitem{Schild} A. Schild, {\it Phys. Rev. } {\bf 131} 2762 (1963);  A. Schild, Science {\bf 138} 994 (1962).  

\bibitem{Schonberg} M. Schonberg, {\it
Phys. Rev.} {\bf 69}, 211 (1946). 

\bibitem{Anderson} J. L. Anderson,  {\em Principles of Relativity Physics },
Academic press, New York (1967), page 225.


\bibitem{Page} L. Page, {\it  Physical Review } {\bf 24}, 296 (1924).  

\bibitem{PRL} J. De Luca, {\it Phys. Rev. Lett.} {\bf 80}, 680 (1998) .

\bibitem{Hans} C. M. Andersen and H. C. von Baeyer, {\it Phys. Rev.
D} {\bf 5}, 802  (1972).

\bibitem{Bellman} R. E. Bellman and K.L.Cooke, Differential-Difference Equations, 
 Academic Press, New York  (1963), page 393.


\bibitem{astar2B} A. Staruszkiewicz, {\it  Acta Physica Polonica}, {\bf XXXIII}, 1007 (1968).


\bibitem{FeyWhe}J. A. Wheeler and R. P. Feynman, {\it Rev. Mod. Phys.} {\bf 17}, 157 (1945); 
J. A. Wheeler and R. P. Feynman, {\it Rev. Mod. Phys.} {\bf 21}, 425 (1949).

\bibitem{Dodd}J. N. Dodd, {\it Eur. J. Phys.} {\bf 4}, 205 (1983).

\bibitem{Hestenes} D. Hestenes, {\it Am. J. Phys.} {\bf 47}, 399 (1979), 
D. Hestenes, {\it Foundations of Physics}  {\bf 15}, 63 (1983). 


\bibitem{Grasman} J. Grasman, {\em Asymptotic Methods for Relaxation Oscillations and Applications}, 
{\it  Applied Mathematical Sciences }, {\bf  63}, Springer-Verlag, New-York (1987). 

\bibitem{to be published} J. De Luca, to be published.


\bibitem{Rohrlich}F. Rohrlich, {\em Classical Charged Particles} Addison-Wesley Publishing, NY (1965).

\bibitem{discrete} J. De Luca, {\it Phys. Rev. E} {\bf 62}, 2060 (2000).

\bibitem{vereador}J.  De Luca , R. Napolitano and V. Bagnato, {\it Physical Review A}, {\bf 55} R1597 (1997), 
J. De Luca, R. Napolitano and  V. Bagnato, {\it Physics Letters A} {\bf 233} 79 (1997).

\bibitem{Mallet-Paret}J. Mallet-Paret, {\it Journal of Dynamics and Differential Equations} {\bf 11}, 1 (1999).


\bibitem{Michael} W. Appel and M. K.-H.Kiessling, {\it Annals of Physics (NY)} {\bf 289}, 24 (2001)


\bibitem{Tomonaga} Sin-itiro Tomonaga, Translated by Takeshi Oka, {\em The story of spin} 
The University of Chicago Press, Ltd., London (1997).

\end{references}
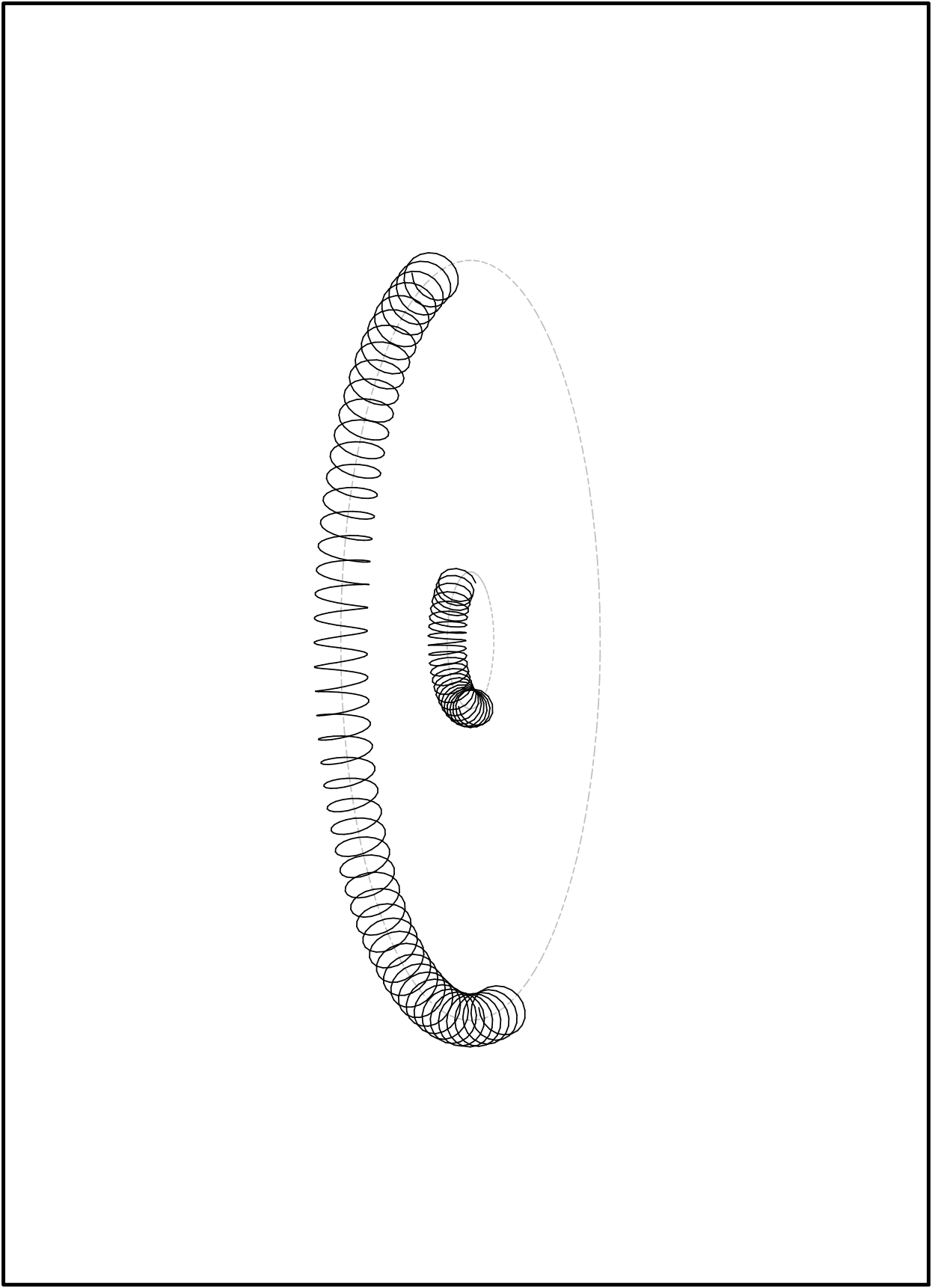
\end{document}